\documentclass[10pt,a4paper]{iopart}
\usepackage[utf8]{inputenc}
\usepackage[english]{babel}
\usepackage[T1]{fontenc}
\usepackage{graphicx}
\usepackage{dcolumn}
\usepackage{bm}
\usepackage[dvipsnames]{xcolor}
\usepackage{tikz}
\usepackage{orcidlink}

\usepackage[bb=dsserif]{mathalpha}

\expandafter\let\csname equation*\endcsname\relax
\expandafter\let\csname endequation*\endcsname\relax
\usepackage{amsmath}

\let\originalleft\left
\let\originalright\right
\renewcommand{\left}{\mathopen{}\mathclose\bgroup\originalleft}
\renewcommand{\right}{\aftergroup\egroup\originalright}

\usepackage[normalem]{ulem}

\usepackage[numbers,sort&compress]{natbib}
\usepackage{doi}

\colorlet{mylinkcolor}{teal}
\colorlet{mycitecolor}{teal}
\colorlet{myurlcolor}{teal}
\usepackage{hyperref}
\hypersetup{
  linkcolor  = mylinkcolor,
  citecolor  = mycitecolor,
  urlcolor   = myurlcolor,
  colorlinks = true,
  breaklinks = true
}

\newcommand{\me}[1]{\textcolor{black}{#1}}
\newcommand{\rr}{\mathcal{R}(\rho_{\bm \lambda})}
\newcommand{\rh}{\mathcal{R}^H(\rho_{\bm \lambda})}
\newcommand{\roh}{\overline{\mathcal{R}}^H(\rho_{\bm \lambda})}
\newcommand{\rhol}{\rho_{\bm \lambda}}
\newcommand{\ket}[1]{| #1 \rangle}
\newcommand{\blambda}{{\bm \lambda}}
\begin{document}

\title[Simultaneous optical phase and loss estimation revisited:  measurement and probe incompatibility]{Simultaneous optical phase and loss estimation revisited:  measurement and probe incompatibility}

\author{Matheus Eiji Ohno Bezerra \footnote{Author to whom any correspondence should be addressed}\,\orcidlink{0000-0003-1550-2575}}
\address{Center for Natural and Human Sciences, Universidade Federal do ABC, 09210-170 Santo Andr\'e, Brazil }
\ead{matheus.eiji@ufabc.edu.br}

\author{Francesco Albarelli\,\orcidlink{0000-0001-5775-168X}}
\address{Scuola Normale Superiore, I-56126 Pisa, Italy}

\author{Rafał Demkowicz-Dobrzanski\,\orcidlink{0000-0001-5550-4431}}
\address{Faculty of Physics, University of Warsaw, Pasteura 5, 02-093 Warsaw, Poland}

\vspace{10pt}
\begin{indented}
\item[] \today
\end{indented}

\begin{abstract}
Quantum multiparameter metrology is hindered by incompatibility issues, such as finding a single probe state (probe incompatibility) and a single measurement (measurement incompatibility) optimal for all parameters. The simultaneous estimation of phase shift and loss in a single optical mode is a paradigmatic multiparameter metrological problem in which such tradeoffs are present. We consider two settings: single-mode or two-mode probes (with a reference lossless mode), and for each setting we consider either Gaussian states or arbitrary quantum states of light restricted only by a maximal number of photons allowed. We find numerically that, as the number of photons increases, there are quantum states of light for which probe incompatibility disappears both in the single- and two-mode scenarios. On the other hand, for Gaussian states, probe incompatibility is present in the single-mode case and may be removed only in the two-mode setting thanks to the entanglement with the reference mode. Finally, we provide strong arguments that the fundamental incompatibility aspect of the model is measurement incompatibility, which persists for all the scenarios considered, and unlike probe-incompatibility cannot be overcome even in the large photon number limit.
\end{abstract}

%
%

\section{\label{intro} Introduction}

Among the many applications of quantum metrology, optical ones are perhaps the most spectacular, as witnessed by the use of squeezed light in gravitational wave interferometric detectors \cite{Schnabel2016, LIGO2019, Virgo2019, LIGO2023}. 
In such a quantum optical setting, photon losses are ubiquitous and represent one of the main limiting factors to quantum-enhanced performance of the devices. 
This is true in a metrological context, where photon loss set a fundamental limitation for the attainable precision in phase estimation~\cite{Escher2011, Demkowicz2012, Demkowicz-Dobrzanski2015a}, but also in other quantum information processing tasks, e.g. quantum communication \cite{Giovanneetti2004}, quantum cryptography \cite{Takeoka2014} or optical quantum computing \cite{Zhong2020}. 
However, estimation of losses themselves is an important problem, related to, e.g. absorption imaging~\cite{Taylor2016} and spectroscopy~\cite{Shi2020}.
Indeed, phase and loss may be considered the two main physical parameters on which information is encoded in the context of optical metrology~\cite{Polino2020}; therefore one would like to optimally estimate both parameters simultaneously with high precision.

This problem falls under the realm of multiparameter quantum metrology, a field concerned with the ultimate precision limits and the optimal strategies for simultaneous estimation of multiple parameters encoded on a quantum system.
This problem is of paramount importance in quantum science, both from a practical and a fundamental perspective, as attested by a growing body of work in the last decade~\cite{Albarelli2019c,Demkowicz-Dobrzanski2020,Pezze2025}.
Multiparameter problems are complicated by the fact that a single probe state and a measurement may be optimal for one parameter but suboptimal for the others; we name these two phenomena probe and measurement incompatibility~\cite{Ragy2016,Albarelli2022}.
As a result, tradeoffs appear in choosing the optimal strategy, since the precision of all parameters cannot be optimized simultaneously.

Measurement incompatibility is an inherently quantum mechanical feature in multiple-parameter estimation, and it appears when considering a parametric family of \emph{quantum states}.
This aspect of the theory has attracted interest from the early days~\cite{Yuen1973}.
The asymptotic scenario in which many copies can be measured collectively is well understood; the fundamental attainable bound is the Holevo Cramér-Rao bound (HCRB)~\cite{Demkowicz-Dobrzanski2020}.
More recent developments focus on bounds tighter than the HCRB for measurements on individual copies~\cite{Conlon2020,Lu2020a,Chen2022f,Chen2024j,Hayashi2023a}.

In a typical metrological scenario, however, the parameter encoding is assumed to be given in the form of a parametric family of \emph{quantum channels}, so that optimization over probe states (or more general probing strategies) becomes crucial.
In the single-parameter case, both for one or many uses of the channel, the effective optimization methods as well as fundamental bounds are well understood~\cite{Zhou2020,Kurdzialek2023a}, but equally powerful multiparameter methods, that would fully grasp all the incompatibility aspects of the problem, are not yet fully developed.
Relevant results, include: bounds to study only probe incompatibility, even for many uses of the channel~\cite{Albarelli2022}, and bounds that take into account all sources of incompatibility, but can be computed only for a single use of the channel~\cite{Hayashi2024} (for moderate Hilbert space dimensions).

It is well known that these incompatibility phenomena are indeed present in the problem of
optical phase and loss estimation~\cite{Crowley2014}. 
Nevertheless, a deeper understanding on the nature of this incompatibility (whether it is the probe- or measurement-incompatibility), as well as the role of the character of the states of light that are considered (whether one is restricted to the use of Gaussian states or one may employ arbitrary states of light limited only by the maximal number of photons present) is missing.

The main goal of this paper is to provide such a insight by collecting some of the scattered results on the problem already present in the literature, as well as contributing new results that fill the gaps. 
Note that we focus on the truly multiparameter version of the problem, where the goal is to obtain information on both parameters simultaneously, and not only 
on some specific combination of the two parameters, which makes the problem effectively a single parameter problem with no fundamental incompatibilty present ~\cite{Dinani2016,Birchall2019}. The summary of the results discussed in the paper is given below.

\begin{figure}[h]
\includegraphics[width=1. \columnwidth]{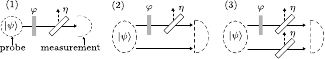}
\caption{\label{fig:3schemes} Three conceptually different simultaneous phase and loss estimation schemes: (1) single-mode scheme, (2) two-mode scheme, where the second mode is not affected by loss, plays the role of a reference mode, and may offer an improved performance thanks to the use of two-mode entangled states of light,
(3) two-mode scheme where loss affects both modes equally. }  
\end{figure}

We consider three conceptually different scenarios of simultaneous phase and loss estimation that are depicted in Fig.~\ref{fig:3schemes}.  
The scenarios differ by the presence or absence of an additional reference mode, and whether the reference mode is also affected by loss.
The choice of a given setting will have a critical impact on the fundamental character of incompatibility---whether it is possible to avoid probe-incompatibility phenomena, measurement-incompatibility phenomena or both.
If both incompatibilities can be avoided there are no fundamental trade-offs between precisions of estimation of the two parameters and both can be estimated simultaneously with the same precision as in the case of strategies optimized independently for each of them~\cite{Ragy2016}. 

The intuition behind the presence of incompatibility in the phase and loss estimation problem comes from realizing that in order to estimate loss it is best to have states with well defined photon number---in fact the optimal state is the $N$-photon Fock state~\cite{Adesso2009}.   
On the other hand, a necessary condition to have large sensitivity to phase is to have large photon number variance in the phase sensing arm~\cite{Escher2011}.

In the most challenging single mode scenario (1) neither measurement nor probe incompatibility can be overcome, when utilizing Gaussian states.
Interestingly, when general single mode states of light are considered (with a cut-off on the maximal number of photons), we find that probe incompatibility appears to vanish
in the asymptotic regime of large photon numbers, thanks to an intricate state structure---this is a numerical confirmation of a previous conjecture based on 
fundamental multiparameter bounds for this problem~\cite{Albarelli2022}.

When a lossless reference beam is allowed (2), obviously one may still avoid probe-incompatibilty issue when using arbitrary states of light, but it is also possible to achieve the same using two-mode entangled Gaussian states as has been pointed out in~\cite{Dowran2021}.
This said, in both scenarios (1) and (2) measurement incompatibility is present for the optimized probe states and hence, in practice one cannot perform simultaneous estimation of phase and loss without trade-offs.

Finally, in scenario (3) where losses affect both modes equally, not only the probe incompatibility issue is not fundamental, both in case of arbitrary and Gaussian states, but so is the measurement incompatibility aspect. This fact has been noted before, in case of arbitrary~\cite{Ragy2016} as well as Gaussian states~\cite{Nichols2018}. We summarize also these results in our presentation for completeness.

The paper is structured as follows. In Section~\ref{preliminaries} we recall the relevant bounds for multiparameter quantum metrology, and we introduce figures of merit to quantify measurement and probe incompatibility. In Section~\ref{stateoftheart} we present all the relevant results on incompatibility issues in phase and loss estimation, and summarize the known results as well as our new findings in Table~\ref{table_results} for all the three schemes considered in Fig.~\ref{fig:3schemes}. In Section~\ref{sec_probe_incomp} we present numerical and analytical results for probe incompatibility for schemes (i) and (ii), both for arbitrary and Gaussian states. In Section~\ref{sec_measurement_incomp} we present numerical and analytical results for measurement incompatibility for schemes (i) and (ii), both for arbitrary and Gaussian states, and also the analysis of paradigmatic optical measurements (photon counting and homodyne). In Section~\ref{sec_conclusions} we offer some concluding remarks.

\section{\label{preliminaries}Multiparameter quantum metrology and incompatibility}

In this section, we first review key concepts from multiparameter quantum estimation theory, then introduce quantifiers for measurement and probe incompatibility.
Here, we keep the presentation general; the specific problem of phase and loss estimation will be discussed in later sections.

\subsection{Multiparameter Cramér-Rao bounds}

The fundamental task in multiparameter quantum metrology is to estimate with the best precision a set of parameters ${\bm \lambda} = \{ \lambda_1,...,\lambda_d \}$ encoded in a probe state $\rho_{\bm \lambda}$, that has been evolved through a quantum channel $\rho_{\bm \lambda} = \Lambda_{\bm \lambda}(\rho)$. 
Some parameters, such as phase, time, and frequency, can only be determined indirectly, as there are no corresponding observables representing these physical quantities.
In order to perform this indirect measurement, we first select a measurement, given by a POVM ${\Pi}_{\bm x}$, which results in measurement outcomes ${\bm x}$ and then choose estimators $\widetilde{\lambda}_i$ that approximate the true parameter value of each parameter $\lambda_i$.

In the multiparameter estimation scenario, the error is quantified by the covariance matrix $\Sigma$, given by
\begin{equation}
    \Sigma_{i j} = \langle \left( \widetilde{\lambda}_i -  \langle\lambda_i \rangle  \right) \left( \widetilde{\lambda}_j -  \langle \lambda_j \rangle \right) \rangle,
\end{equation}
where the diagonal elements are the variances $\Delta^2 \widetilde{\lambda }_i = \Sigma_{i i}$. For any unbiased estimator, the quantum Cramér-Rao bound (CRB) gives a fundamental lower bound to the covariance matrix, which in the multiparameter case reads: 
\begin{equation}
    \Sigma \geq F(\rho_{\bm \lambda})^{-1},
    \label{CRB_matrix}
\end{equation}
where $ F(\rho_{\bm \lambda})$ is the quantum Fisher information (QFI) matrix, defined as
\begin{equation}
    F_{i j} (\rho_{\bm \lambda}) = \frac{1}{2} \Tr \big( \rho_{\bm \lambda} \{ L_{\lambda_i} , L_{\lambda_j} \}  \big)
    \qquad
    \frac{\partial \rho_{\bm \lambda}}{\partial \lambda_i} = \frac{1}{2} \{ \rho_{\bm \lambda} , L_{\lambda_i} \} ,
    \label{qfi}
\end{equation}
where $L_{\lambda_i}$ is the symmetric logarithmic derivative (SLD) related to the parameter $\lambda_i$.
The diagonal elements of Eq.~\eqref{qfi} give the QFI for the corresponding parameter $\lambda_i$, which determines the quantum precision limit for estimating this parameter independently of the others. The off-diagonal elements of the QFI matrix indicate correlations between the estimation of different parameters $\lambda_i$ and $\lambda_j$. We can recast the matricial quantum CRB given by Eq.~\eqref{CRB_matrix} into a scalar quantum CRB as follows:
\begin{equation}
    \Tr \left(W \Sigma \right) \geq C^S(\rho_{\bm \lambda}) = \Tr \left( W F(\rho_{\bm \lambda})^{-1} \right) ,
\end{equation}
where the weight matrix $W$ is a real, positive, $d \times d $ matrix. 

The quantum CRB depends only on the state $\rho_{\bm \lambda}$ and, in general, cannot be saturated for all parameters by considering one single measurement strategy \cite{Holevo2011b,Demkowicz-Dobrzanski2020}.
The optimal measurement strategy for a parameter is determined by its corresponding SLD and then, when the SLDs do not commute, we expect that the bound cannot be attained.
A necessary and sufficient (assuming access to multiple copies of the state) condition for saturating the QCR bound is that the
commutators of SLDs vanish when traced with the state \cite{Ragy2016,Suzuki2018}:
\begin{equation}
    I_{\lambda_i \lambda_j} (\rho_{\bm \lambda}) = \frac{1}{2} \Tr \big( \rho_{\bm \lambda} \left[ L_{\lambda_i} , L_{\lambda_j} \right]  \big).
    \label{incomp}
\end{equation}
Otherwise, if this expectation value is non-zero, it indicates measurement incompatibility, meaning the parameters cannot be estimated simultaneously with the precision given by the CRB. 

In order to take into account the measurements' incompatibilities, a tighter bound is introduced, the Holevo-Cramér-Rao bound (HCRB) \cite{Holevo2011b,Nagaoka1989}, given in terms of the following double minimization problem:
\begin{equation}
 \Tr \left(W \Sigma \right) \geq    C^H (\rho_{\bm \lambda}) = \min_{V \in S^n} \min_{{\bm X} \in \chi_{\bm \lambda}} \left[ \Tr \big(W V) | V \geq Z(X) \right]
 \label{hcrb}
\end{equation}
where  $\left[ Z(X) \right]_{ij} = \Tr \left( X_i X_j \hat{\rho_{\bm \lambda}} \right)$, $S^n$ is the set of $n \times n$ symmetric matrices and 
\begin{equation}
\chi_{\bm \lambda} = \left\{ {\bm X}=(X_1,...,X_n) \hspace{1mm} \Big| \hspace{1mm} \Tr \left( X_i \frac{\partial \rho_{\bm \lambda}}{\partial \lambda_j}   \right) = \delta_{ij}  \right\}
\end{equation}
represent a set of locally unbiased observables, that are used to estimate each of the parameters respectively. 
Then, we have the following chain of inequalities giving the lower bounds for the scalar multiparameter cost:
\begin{equation}
    \Tr \left( W \Sigma \right) \geq C^H(\rhol) \geq C^S(\rhol).
    \label{bounds_CR_HCR}
\end{equation}
Additionally, since the calculation of the HCRB is often challenging, an explicit upper bound for the HCRB can be derived in terms of the SLD operators~\cite{Carollo2019, Albarelli2019c,Tsang2019}, resulting in the following  chain of inequalities: 
\begin{equation}
\label{eq:bounds2}
C^S (\rho_{\bm \lambda}) \leq C^H(\rho_{\bm \lambda}) \leq \overline{C}^H(\rho_{\bm \lambda})  \leq 2 C^S(\rho_{\bm \lambda}),
\end{equation}
where
\begin{equation}
     \overline{C}^H (\rho_{\bm \lambda}) = C^S (\rho_{\bm \lambda}) + \left \Vert \sqrt{W} F(\rho_{\bm \lambda})^{-1} 
    I(\rho_{\bm \lambda}) F(\rho_{\bm \lambda})^{-1} \sqrt{W}  \right\Vert_1 ,
    \label{upper_hcrb}
\end{equation}
and $ \left \Vert  O \right\Vert_1 := \Tr \left( \sqrt{ O^\dagger O } \right)$ is the trace norm.
The last term contains the measurement incompatibility term and vanishes only when the SLDs commutators vanish when traced with the state, resulting then in $I (\rho_{\bm \lambda}) = 0$. Indeed, both bounds are equivalent in D-invariant models~\cite{Albarelli2019c,Demkowicz-Dobrzanski2020}.

\subsection{Measurement incompatibility quantifier}

In this paper we will use the following expression to quantify measurement incompatibility:
\begin{equation}
\rr = \frac{C^S(\rhol)}{\min\Tr(W \Sigma)} \leq 1,
\label{minim_cost}
\end{equation}
where the minimization is performed over all locally unbiased measurements\footnote{Notice that this is equivalent to minimizing the weighted trace of the inverse classical Fisher information matrix over all POVMs.}.
If this quantity equals $1$, this means that the CRB is saturable; otherwise, if it is strictly smaller, it is an indication of a fundamental measurement incompatibility present in the model.

As it may be hard to find optimal measurements that minimize the multiparameter estimation cost, it will be convenient to introduce easier computable measurement incompatibility indicators based on the HCRB:
\begin{equation}
\rh =  \frac{C^S(\rhol)}{C^H(\rhol)}, \quad \roh =  \frac{C^S(\rhol)}{\overline{C}^H(\rhol)}.
\label{r_h}
\end{equation}
From \eqref{bounds_CR_HCR} it follows that $\mathcal{R}(\rhol) \leq \mathcal{R}^H(\rhol)$.
Hence, if we observe that $\rh < 1$, this implies the presence of measurement incompatibility.
Moreover, if $\rh=1$, then in the asymptotic limit $N \rightarrow \infty$ we may use general arguments for asymptotic saturability of the HCRB for models with linear scaling of QFI~\cite{Albarelli2022} and conclude no asymptotic measurement incompatibility.

In the case of D-invariant models $C^H(\rhol) = \overline{C}^H(\rhol)$~\cite{Holevo2011b} and thus $\rh = \roh$, so the analysis is simpler. Nevertheless, also for general models which are not necessarily D-invariant, we may still compute $\roh$ to get a sufficient criterion for measurement compatibility---if $\roh=1$ then it follows from \eqref{eq:bounds2} that
$\mathcal{R}^H(\rhol)=1$ and by asymptotic saturability arguments of HCRB we may conclude that asymptotically $\rr \overset{N \rightarrow \infty}{\rightarrow} 1$.

We stress that the measurement incompatibility quantifiers $\rr, \rh$ and $\roh$ all depend on the choice of the cost weight matrix $W$. 
Alternative figures of merit for quantifying measurement incompatibility, independent of $W$ (i.e., invariant under model reparametrizations), have been introduced~\cite{Carollo2019,Belliardo2021,Candeloro2021c}.
We choose to use the quantities defined above because: i) we care about the physical separation between phase and loss parameters, and ii) we will later choose a specific diagonal $W$ to compare the error in each parameter with the corresponding optimal single-parameter counterpart.

\subsection{Probe incompatibility quantifier}

In multiparameter quantum metrology, a key problem is to determine whether a single state exists that performs optimally when estimating all the parameters simultaneously.
In order to analyze this problem quantitatively, we introduce a figure of merit to quantify the amount of probe incompatibility of a given probe state $\ket{\psi}$.
We consider the sum of the diagonal elements of the QFI matrix, each rescaled by the respective optimal single parameter QFI
\begin{equation}
    \mathcal{F} (\rho_{\bm \lambda}) = \frac{1}{d} \sum_{j=1}^d \frac{F_{j j} (\rho_{\bm \lambda})}{F_{\lambda_j}^{(\text{max})}} \leq 1,
    \label{probe_incomp},
\end{equation}
where $\rho_{\bm \lambda} = \Lambda_{\bm \lambda}(| \psi \rangle \langle \psi | )$ and $F_{\lambda_j}^{(\text{max})}$ is the QFI of the single parameter $\lambda_j$ optimized over input probe states.
This quantity is bounded as $0 \leq \mathcal{F}(\Lambda_{\bm \lambda}) \leq 1$, and the value $1$ indicates no probe incompatibility in the model, meaning that each diagonal element of the QFI matrix attains the value of the optimized single-parameter QFI.

Even if there is no probe incompatibility, there could also be additional issues due to correlations between the estimators of the parameters.
Mathematically, this is related to the fact that the CRB is given by the inverse of the QFI matrix, the correlations are related to the off-diagonal elements, and they imply the inequality $\Tr\left( F(\rho_{\bm \lambda})^{-1} \right) \geq \sum_j 1/F(\rho_{\bm \lambda})_{jj}$. While we will not directly consider the impact of such correlations in optimizing the probe state, it is always possible to check the full QFI matrix afterwards, to see if the off-diagonal elements are sufficiently small.

In order to investigate probe-incompatibility of the problem we, therefore, need to maximize $\mathcal{F}(\rho_{\bm \lambda})$~\eqref{probe_incomp} over input probe states $\ket{\psi}$. As a result we will obtain a quantity that depends only on the channel $\Lambda_{\bm \lambda}$ and captures the intrinsic probe incompatibility of the metrological model.

\subsection{Iterative see-saw optimization of the probe-incompatibility quantifier}
In this paper we will be using an iterative see-saw algorithm (ISS) to maximize $\mathcal{F}(\rho_{\bm \lambda})$ over probe states. This algorithm can be viewed as a direct generalization of the ISS method used in single-parameter quantum metrology~\cite{Macieszczak2014,Macieszczak2013, Kurdzialek2025}, which, however, has not been applied to study multiparameter estimation problems until now.

To start, we recall some details about the single parameter case. In this case, the optimization consists in the maximization of the QFI over the input states, i.e. $\max_{| \psi \rangle} F_{\lambda_i} (\rho_{\bm \lambda})$ with $\rho_{\bm \lambda} = \Lambda_{\bm \lambda}(| \psi \rangle \langle \psi | )$. We introduce the pre-QFI function for the parameter $\lambda_i$, defined as follows:
\begin{equation}
    f_{\lambda_i} (| \psi \rangle, A) = 2 \hspace{1mm} \Tr \left(\frac{\partial \rho_{\bm \lambda}}{\partial \lambda_i} A \right) - \Tr \left( \rho_{\bm \lambda} A^2 \right)
    \label{pre_qfi}
\end{equation}
where $A$ is a Hermitian operator.
A key point is that maximizing the previous function over the operator $A$ results in the QFI of this corresponding parameter, where the optimal $A$ is the corresponding SLD $L_{\lambda_i}$ given in Eq.~\eqref{qfi}.
Then, the ISS algorithm consists in the double maximization problem:  $\max_{| \psi \rangle} F_{\lambda_i} (\rho_{\bm \lambda}) = \max_{| \psi \rangle } \max_L f_{\lambda_i} (| \psi \rangle , A)$. Indeed, in the step of maximizing over $| \psi \rangle$, we first rewrite the pre-QFI function in terms of the dual map of the quantum channel:
\begin{equation}
    f_{\lambda_i} (| \psi \rangle, A) = \Tr \Big( | \psi \rangle 
    \langle \psi | M_{\lambda_i} \Big)
    \qquad
    M_{\lambda_i} = 2 \frac{ \partial \Lambda^*}{\partial \lambda_i} (A) - \Lambda^* (A^2) 
    \label{qfi_algorithm}
\end{equation}
where $\Lambda^*(\cdot)=\sum_m K_m^\dagger \cdot K_m$, resulting in the optimal $| \psi \rangle$ being the eigenvector with the largest eigenvalue of the matrix $M_{\lambda_i}$.
The practical implementation of this algorithm is already discussed in previous works~\cite{Macieszczak2014,Macieszczak2013, Kurdzialek2025}.

A generalization of this method for the case of multiple parameters, can be obtained by introducing the following multiparameter pre-QFI:
\begin{equation}
    f_{\bm \lambda} (| \psi \rangle, \{ A_{\lambda_j} \}_{j=1}^d ) = \sum_{j=1}^d \frac{f_{\lambda_j} (| \psi \rangle, A_{\lambda_j} )}{\omega_{\lambda_j}} 
\end{equation}
where each $f_{\lambda_i} (| \psi \rangle, A_{\lambda_i})$ is defined as in Eq.~\eqref{pre_qfi} and $\omega_{\lambda_j} > 0 $ are generic positive weights.
These are chosen as $\omega_{\lambda_j}= F_{\lambda_j}^{(\text{max})} $ to compute the probe incompatibility quantifier~\eqref{probe_incomp}.
Then, the iterative algorithm for simultaneous estimation works in the following way: [i] Given a random input state $| \psi \rangle$, the first step is the maximization of $f_{\bm \lambda} (| \psi \rangle, \{ A_{\lambda_j} \}_{j=1}^d )$ over the $d$ hermitian operators $A_{\lambda_j}$ for $j=1,\dots, d$, obtaining the corresponding SLDs $L_{\lambda_j}$. 
[ii] For the second step, we plug the previous result in each term of Eq.~\eqref{qfi_algorithm} and perform the maximization:
\begin{equation}
    \max_{| \psi \rangle} \left( \sum_{j=1}^d \frac{f_{\lambda_j} (| \psi \rangle, A_{\lambda_j} )}{\omega_{\lambda_j}} \right) = \max_{| \psi \rangle} \Tr \left( | \psi \rangle \langle \psi | M\right) 
    \qquad
    M =  \sum_{j=1}^d \frac{M_{\lambda_j}}{\omega_{\lambda_j}}
    \label{normalized_m_matrix}
\end{equation}
which results in the optimal state $| \psi \rangle$ being the eigenvector with the largest eigenvalue of the matrix $M$, denoted by $| \psi^{[ii]} \rangle$. Then, we repeat this iteration until the algorithm converges (e.g., when the last five results do not differ by more than $0.1 \%$).

\subsection{Necessary conditions for probe-compatibility from asymptotic upper bounds}
\label{sec:probeincompbound}
The ISS optimization described above may be viewed as a way to obtain a lower bound on $\mathcal{F}(\rhol)$.
On the other hand, we may take a complementary approach and try to upper bound $\mathcal{F}(\rhol)$ using fundamental upper bounds on the weighted sum of QFIs in general multiparameter estimation problems, found by minimization over different Kraus representations of a quantum channel~\cite{Albarelli2022}.

Given a quantum channel $\Lambda_{\bm \lambda}$ its Kraus representation 
\begin{equation}
\Lambda_{\bm \lambda}(\rho) = \sum_m K_{{\bm \lambda}, m} \rho K_{{\bm \lambda}, m}^\dagger   
\end{equation}
is not unique, and equivalent Kraus representations are connected with each other via a unitary matrix (more generally an isometry)~\cite{Nielsen2000}
\begin{equation}
\label{eq:Kraus_equiv}
    \tilde{K}_{{\bm \lambda}, m} = \sum_{m^\prime} u(\bm \lambda)_{m}^{m^\prime} K_{{\bm \lambda},m^\prime},
\end{equation}
where $u({\bm \lambda})$ is a unitary matrix, which importantly may depend on the estimated parameters.
In what follows we will drop explicit dependence of Kraus operators on $\blambda$ for conciseness.

In the single parameter case, one can show that the maximal achievable QFI for the output state of the channel, optimized over all input probe states, is upper bounded by~\cite{Fujiwara2008,Escher2011,Demkowicz2012}:
\begin{equation}
    \max_{\rho}F\left[\Lambda_{\lambda}(\rho)\right] \leq 4 \min_{\{K_{m}\}} \left\| \sum_m \partial_\lambda K_{m}^\dagger \partial_\lambda K_{m}\right\|,
\end{equation}
where minimization is performed over all equivalent Kraus representations of the channel and $\| \cdot \|$ is the operator norm.\footnote{The inequality becomes in fact equality, if one admits the possibility that the probe system may be entangled with a noiseless ancillary system on which the channel acts trivially.}
For finite-dimensional systems the above minimization may be performed efficiently, and cast in the form of a semidefinite program (SDP)~\cite{Demkowicz2012}.
Nevertheless, the bound is valid even if we do not perform full minimization over all Kraus representations and consider just a certain subclass. 

The above bound has been generalized in~\cite{Albarelli2022} to the multiparameter case, in order to upper bound the weighted sum of QFIs\footnote{\me{Similarly to the single-parameter case, if the probe state is allowed to be entangled with a noiseless ancillary system, this bound is saturable.}}, and hence it may be applied to upper bound the probe-incompatiblity measure 
\eqref{probe_incomp}:
\begin{equation}
\label{eq:probeincompbound}
\max_{\rho} \mathcal{F}[\Lambda_{\bm \lambda}(\rho)] \leq \frac{4}{d} \min_{\{K_{\lambda,m}\}}  \left\| \sum_{j=1}^d \frac{1}{ w_{\lambda_j}} \sum_{m} \partial_{\lambda_j} K_{m}^\dagger \partial_{\lambda_j} K_{m} \right\|, \ w_{\lambda_j} = F_{\lambda_j}^{\mathrm{(max)}},
\end{equation}
where $w_{\lambda_j}$ play the role of the weights.
Thus, if $\max_\rho \mathcal{F}[\Lambda_{\bm \lambda}(\rho)] < 1$ the model suffers from fundamental probe-incompatibility.
In contrast, the bound being equal to $1$ is a necessary condition for probe compatibility in the model~\cite{Albarelli2022}.

The above bound is expressed in terms of operator norms, well suited when we deal with small finite-dimensional spaces.
In the case of large or infinite-dimensional spaces, we usually impose some additional constraints on the states that are allowed in the problem.
In the case of optical interferometry, this usually amounts to a restriction on the mean photon number.
In such scenarios, it is more convenient to rewrite the above bounds, replacing the operator norms with expectation values on the states that are allowed in the problem.
For single-parameter bounds, this approach has been followed in~\cite{Escher2011}.\footnote{In \cite{Escher2011}, the bound includes also a second subtracted term, which, however, is irrelevant, as one may always choose such a Kraus representation, that the second term vanishes, see e.g. the discussion in \cite{KolodynskiPhd2014}, Appendix C.}
One can do the same in a straightforward way in case of multiparameter bound \eqref{eq:probeincompbound} and write:
\begin{equation}
\label{eq:probeincompboundexp}
\max_{\rho \in \mathcal{S}} \mathcal{F}[\Lambda_{\bm \lambda}(\rho)] \leq 
 \frac{4}{d} \min_{\{K_{m}\}} \max_{\rho \in \mathcal{S}}  \Tr \left[ \rho \sum_{j=1}^d \frac{1}{w_{\lambda_j}} \sum_{m} \partial_{\lambda_j} K_{,m}^\dagger \partial_{\lambda_j} K_{m} \right],
\end{equation}
where we restrict the set of input states to some subset $\mathcal{S}$. In Sec.~\ref{sec:necprobecomp} we will use this bound to obtain necessary conditions for probe-compatibility in simultaneous phase and loss estimation, expressed in terms of photon-number statistics properties that the state needs to satisfy.

\vspace{2mm}

\section{\label{stateoftheart}State of the art and summary of the results}

\subsection{Ultimate metrological limits for independent phase and loss estimation}

Let us now restrict our attention to phase and loss estimation, so a two-parameter estimation problem where ${\bm \lambda} = \{\varphi, \eta \}$, and $\Lambda_{\bm \lambda}$ represents the combined action of phase delay and loss.
If one treats the problem of phase and loss independently, then optimal protocols that saturate fundamental bounds are well known.

In optical phase estimation, a coherent state achieves the standard quantum limit (SQL) scaling, whereas interfering with a single-mode squeezed state enhances the precision, achieving a sub-SQL scaling~\cite{Caves1981}.
It was proven that both a NOON state~\cite{Bollinger1996} and general two-mode squeezed state~\cite{Anisimov2010} can achieve the Heisenberg scaling for phase estimation.
However, their performance is highly susceptible to photon losses, which rapidly degrade that advantage.
Phase estimation in the presence of loss was first analyzed in~\cite{Demkowicz-Dobrzanski2009, Dorner2009}, where it was numerically shown that the optimal states exhibit a non-trivial structure as a
function of the loss.
Going beyond the analysis of particular strategies, it is possible to obtain the ultimate bounds on the maximal QFI for phase estimation in the presence of loss, which is valid for any probe state with maximal $N$ photons and moreover is saturable in the asymptotic limit $N \rightarrow \infty$~\cite{Koodynski2010, Knysh2011, Escher2011, Demkowicz2012}:
\begin{equation}
    F^{\text{(max),(1,2)}}_\varphi \overset{N \rightarrow \infty}{\longrightarrow} \frac{4 \eta N}{1-\eta}, \quad F^{\text{(max),(3)}}_\varphi \overset{N \rightarrow \infty}{\longrightarrow} \frac{\eta N}{1-\eta} 
    \label{upper_phase},
\end{equation}
where $F^{\text{(max),(1,2)}}_\varphi$ represents the maximal achievable QFI in scenarios (1), (2) where loss affects only a single mode, and (3) in case where both modes are affected equally. 
Importantly, these bounds can be asymptotically saturated via interferometry schemes involving squeezed states~\cite{Demkowicz-Dobrzanski2013} (in this case $N$ should be interpreted as the mean number of photons used).

Analogously, the fundamental bound for the estimation of the loss parameter is~\cite{Monras2007,Adesso2009,Nair2018}
\begin{equation}
    F^{\text{(max)}}_\eta = \frac{N}{\eta(1 -\eta)},
    \label{upper_loss}
\end{equation}
it is achievable by a Fock state \cite{Adesso2009} (for any finite $N$) or, interpreting $N$ as the mean photon number, by any pure state diagonal in the Fock basis~\cite{Nair2018}, which includes two-mode Gaussian states.

We also mention that the phase and loss ultimate limits in Eqs.~\eqref{upper_phase} and~\eqref{upper_loss}, which are achieved by very different states, can also be related by realizing that a phase profile at various frequencies must be accompanied by a loss profile, according to the Kramers-Kronig relations~\cite{Gianani2021}.

In the multiparameter setting, a key problem is to determine whether the above asymptotic bounds, which are saturable in single parameter scenarios, can be  saturated when phase and loss are being sensed simultaneously.
In order to analyze the problem quantitatively, we will
use the two measures for probe and measurement incompatibility measures, respectively $\mathcal{F}(\rho_{\bm \lambda})$ and $\rr$ that we have defined for general models in Sec.~\ref{preliminaries}.

Table~\ref{table_results} provides a concise summary of the incompatibility aspects of the three models discussed in this paper (depicted in Fig.~\ref{fig:3schemes}), indicating the behavior of these two quantities (or the lower bound on $\roh$ in lieu of $\rr$). For the measurement incompatibility indicators, we choose the weight matrix $W=\text{diag}(F_\varphi^{(\text{max})}, F_\eta^{(\text{max})})$, since it produces asymptotically a more regularized bound, similarly to the probe incompatibility figure of merit~\eqref{probe_incomp}.

\begin{table}
    \centering
\begin{tabular}{||c c c||} 
     \hline
      & Optimal $N$-photon states & Gaussian states \\ [0.5ex] 
     \hline\hline
     (1) Single-mode & \begin{tabular}{c} { Probe compatibility} \\  $\mathcal{F} (\rho_{\bm \lambda})
     \rightarrow 1 $  \\
     Meas. incompatibility: \\  $\roh \overset{\eta \neq 1}{\rightarrow} \frac{1}{2}$ 
      \end{tabular} & \begin{tabular}{c} Probe incompatibility: \\ $\mathcal{F} (\rho_{\bm \lambda}) \rightarrow \frac{1}{2} $ \end{tabular}  \\ 
     \hline
     \begin{tabular}{c} (2) Two-modes\\
     (single mode loss)\\
     \end{tabular} & \begin{tabular}{c} Probe compatibility \cite{Albarelli2022}: \\ $\mathcal{F} (\rho_{\bm \lambda}) \rightarrow 1 $ \\ Meas. incompatibility \cite{Crowley2014}: \\ $\roh \overset{\eta \neq 1}{\rightarrow} \frac{2}{3}$  \end{tabular} & \begin{tabular}{c} Probe compatibility: \\ $\mathcal{F} (\rho_{\bm \lambda}) \overset{\bar{N}_\alpha \gg \bar{N}_r}{\rightarrow} 1 $ \\Meas. incompatibility: \\ $ \roh \overset{\eta \neq 1}{\rightarrow} \frac{1}2$ \end{tabular} \\
     \hline
     \begin{tabular}{c} (3) Two-modes\\
     (loss in both modes)
     \end{tabular} & \begin{tabular}{c} Probe compatibility \cite{Ragy2016}: \\ $\mathcal{F} (\rho_{\bm \lambda}) = 1 $ \\  Meas. compatibility \cite{Ragy2016}: \\ $ \rr = 1$ \end{tabular}  & \begin{tabular}{c} Probe compatibility \cite{Nichols2018}: \\ $\mathcal{F} (\rho_{\bm \lambda}) \rightarrow 1 $  \\  Meas. compatibility \cite{Nichols2018}: \\ $ \rr \rightarrow 1$ \end{tabular} 
     \\
     \hline
     \hline
    \end{tabular}
    \caption{Summary measurement and probe incompatibility in simultaneous phase and loss estimation, including both new and previously known results.
    The figure of merit $\mathcal{F}(\rho_{\bm \lambda})$ is defined in Eq.~\eqref{probe_incomp}, while $\rr$ and $\roh$ in Eqs.~~\eqref{measurement_incomp} and \eqref{r_h}} 
    \label{table_results}
\end{table}

\subsection{Compatibility in case of two-mode losses scenario (3)}

As indicated in Table~\ref{table_results}, the two-mode (loss in both modes) scenario (3) is distinct form the others as there is no fundamental incompatibility present in this case neither in terms of probe nor measurement. Here we will briefly explain the reasons behind this phenomena, while in the rest of the paper we will solely focus on scenarios  (1) and (2) where incompatibility aspects affect the achievable estimation precision in a non-trivial way.

Consider a protocol involving some two-mode $N$-photon state:
\begin{equation}
    \ket{\psi_N^{(3)}} = \sum_{n=0}^N c_n^{(3)} \ket{n} \ket{N-n},
\end{equation}
where $\ket{n}\ket{N-n}$ represents a state where $n$ photons go the upper and $N-n$ the lower arm respectively, and superposition coefficients $c_n^{(3)}$ are chosen in such a way that phase estimation precision achieves the fundamental bound $F^{(\text{max}), (3)}$ as given in \eqref{upper_phase}. Note that, in all terms of the superposition there is the same total number of photons $N$. Since losses are equal in both modes, the way the photons are split between the modes does not matter, and from the point of view of estimating losses, this state behaves equally well as a Fock state with $N$ photons.
Hence, there is no probe-incompatibility in this scenario. Moreover, in order to estimate loss optimally it is enough to  measure the total number of photons that make it through. This is also compatible with optimal measurements that are used to estimate the phase, which involve interfering the output modes and measuring photons in respective output ports of the interferometer---this operation does not change the total number of photons measured in both modes \cite{Ragy2016}.

Moreover, we can also avoid any incompatibility issues if we are restricted to the use of Gaussian states~\cite{Nichols2018}.
In this case, one has to be more careful with the choice of Gaussian states, as not all the states that are optimal for phase estimation will be automatically optimal for loss estimation, due to potentially too large photon number fluctuations.
To begin, we define a displacement operator acting in the mode $k$ is as follows:
\begin{equation}
    D^{(k)}(\alpha_k) = \text{exp}\left( \alpha_k \text{e}^{i \mu_k} \hat{a}^\dagger_k -  \alpha_k \text{e}^{-i \mu_k} \hat{a}_k \right) ,
    \label{displacement}
\end{equation}
where $\bar{N}_\alpha = |\alpha|^2$ gives the average photon number generated and $\mu$ denotes the direction of the displacement in the phase space.
We also introduce the two-mode squeezing operator as follows:
\begin{equation}
    S^{(2)}(r) 
     = \text{exp} \Big( r  \text{e}^{-i \theta} \hat{a}_1 \hat{a}_2 - r  \text{e}^{i \theta} \hat{a}^\dagger_1 \hat{a}^\dagger_2 \Big),
\end{equation}
where $\bar{N}_r = \hspace{1mm} \sinh^2 r$ gives the average photon number generated and $\theta$ denotes the direction of the squeezing in the phase space.
Choosing a displaced two-mode squeezed probe state, defined by:
\begin{equation}
    \ket{\psi_G^{(3)}}  = D^{(1)}(\alpha_1) D^{(2)}(\alpha_2) S^{(2)}(r) | 0, 0 \rangle ,
\end{equation}
it was shown in~\cite{Nichols2018} that when $|\alpha_1|=|\alpha_2|$ and $\mu_1 = \mu_2 = i$, this state can overcome the measurement incompatibility condition given by Eq.~\eqref{incomp}.
Indeed, in the asymptotic limit, the regime with strong displacement is optimal for the simultaneous estimation of phase, loss and thermal noise.
Furthermore, based on the explicit expression of the QFI matrix in the asymptotic approximation given in~\cite{Nichols2018}, this state can asymptotically saturate the optimal QFI bounds for phase and loss estimation, given by Eqs.~\eqref{upper_phase} and~\eqref{upper_loss}, in the absence of thermal noise, even though this connection was not highlighted in~\cite{Nichols2018}.

\section{Probe incompatibility for single-mode loss}
\label{sec_probe_incomp}
\begin{figure}[t]
    \centering
\includegraphics[width=0.75 \columnwidth]{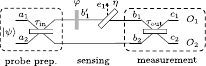}
\caption{\label{scheme} Interferommetric scheme considered. \textit{Probe preparation:} Pure state $| \psi \rangle$ enters the interferometer, goes through a beamsplitter with transmissivity $\tau_{\mathrm{in}}$. \textit{Parameters encoding:} In sequence, a phase shift $\varphi$ is applied, resulting in a unitary evolution $\{ \hat{a}_1 , \hat{a}_2 \} \rightarrow \{ \hat{b}'_1 , \hat{b}_2 \}$; followed by the effect of loss, modelled by an evolution $b'_1 \rightarrow \{\hat{b}_1 , \hat{c}_1 \}$. This stage is discussed in Sec.\ref{sec_probe_incomp}, where we investigate the fundamental quantum precision limits and the probe incompatibilities, \textit{Measurement:} The output state is transformed by the action of an output beamsplitter, leading the evolution $\{ \hat{b}_1 , \hat{b}_2 \} \rightarrow \{ \hat{c}_1 , \hat{c}_2 \}$ and then subjected to a measurement strategy; photon counting and homodyne detection. This stage is discussed in Sec.~\ref{sec_measurement_incomp},  where we investigate the precision limits including the measurement incompatibilities.}
\end{figure}

We are now ready to focus on scenarios (1) and  (2). 
In order to have better physical intuitions for optimal protocols, especially in the case of Gaussian states, we will consider an interferometric scheme where the initial and final beam splitters are explicitly included in the description and have tunable transmissivities, $\tau_{\text{in}}$, $\tau_{\text{out}}$ as depicted in Fig.~\ref{scheme}. 
The loss affects only the upper arm, while the second arm serves as a reference in case of scenario (2).

We denote by $\hat{a}_k$ ($\hat{b}_k$) the bosonic annihilation operators acting in the input (output) mode $k$. First, we have the unitary evolution $\{ \hat{a}_1 , \hat{a}_2 \} \rightarrow \{ \hat{b}'_1 , \hat{b}_2 \}$ due to the action of an input beamsplitter with transmissivity $\tau_{\mathrm{in}} \in [0, 1]$ followed by the phase shift $\varphi \in [0, 2 \pi]$, given the following evolution:
\begin{equation}
    \begin{pmatrix}
    \hat{b}'_1 \\
    \hat{b}_2
    \end{pmatrix}
    =
    U
    \begin{pmatrix}
    \hat{a}_1 \\
    \hat{a}_2
    \end{pmatrix}
    \qquad
    U = 
    \begin{pmatrix}
    \e^{i \varphi} & 0  \\
    0 & 1
    \end{pmatrix}
    \begin{pmatrix}
    \sqrt{\tau_{\mathrm{in}}} & i \sqrt{1-\tau_{\mathrm{in}}}  \\
    i \sqrt{1-\tau_{\mathrm{in}}} & \sqrt{\tau_{\mathrm{in}}}
    \end{pmatrix}
    .
    \label{unitary_evol}
\end{equation} 
Subsequently, the effects of the loss are modeled via the action of a fictional beamsplitter with transmissivity $ \eta \in [ 0 , 1 ]$ as follows: 
\begin{equation}
    \hat{b}_1 = \sqrt{\eta} \hspace{0.5mm} \hat{b}'_1 + \sqrt{1-\eta} \hspace{0.5mm} \hat{e}_1 ,
    \label{non_unitary_evol}
\end{equation}
where $\hat{e}_1$ is a virtual mode where the photons are lost and which we do not have access to.
In that way, given a probe state (pure) $| \psi \rangle$ at the input, its evolution is given by  Eqs.~\eqref{unitary_evol}, \eqref{non_unitary_evol}, resulting in a mixed state at the output, denoted as $\rho_{\bm \lambda} = \Lambda_{\bm \lambda}(| \psi \rangle \langle \psi | )$, where $\bm \lambda = (\varphi, \eta)$.

In this section, the main objective is to analyze the probe incompatibility in the simultaneous estimation of phase and loss, by looking at the maximal achievable value of normalized total QFI $\mathcal{F}(\rhol)$ defined in \eqref{probe_incomp}.
Recall that $\mathcal{F}(\rhol)=1$ represents no probe-incompatibility case, while  $1/2$ means that half of the photons are used to estimate one parameter and half to estimate the other---for our problem involving two parameters, this is the strongest probe-incompatibility case. 

We will first make use of the fundamental bound discussed in Sec.~\ref{sec:probeincompbound} to provide necessary conditions the state needs to satisfy in order to achieve probe-compatiblity in simultaneous loss and phase estimation, and then proceed to search for the optimal states. 

\subsection{Necessary conditions for probe-compatibility in phase and loss estimation}
\label{sec:necprobecomp}

First, note that the bounds in Sec.~\ref{sec:probeincompbound} remain valid even when noiseless ancillary systems entangled with the probe are used, i.e. when the channel acts trivially on the ancillas.
This makes the bounds insensitive to the distinction between scenarios (1) and (2) in Fig.~\ref{fig:3schemes}.
We will, therefore, focus on the simplest single mode scenario (1), and denote the annihilation operators for this mode as $\hat{b}$ for conciseness. 

Following~\cite{Escher2011}, let us consider the following Kraus representation of the single mode channel $\Lambda_{\bm \lambda}$, ${\bm \lambda} =(\varphi, \eta)$
\begin{equation}
    \label{eq:KrausOps_alpha_beta}
K_{ m} = e^{-i \varphi \beta}\sqrt{\frac{(1-\eta)^m}{m!}}e^{i \varphi (\hat{n} - \alpha m)}\eta^{\frac{n}{2}} \hat{b}^m, \quad m=0,1,\dots, 
\end{equation}
where $\hat{n}=\hat{b}^\dagger \hat{b}$ is the photon number operator, while $\alpha$, $\beta$ are free real parameters that are used to obtain different equivalent Kraus representations.\footnote{Compared with \cite{Escher2011}, we introduced additional $\beta$ parameter, which is needed to compensate for the lack of additional subtracted term in the bound we are using compared to the one in \cite{Escher2011}.}
A Kraus operator $K_{m}$ represents an event where $m$ photons are lost from the mode.  

After some algebraic calculations, that make extensive use of commutation properties of annihilation operators, one arrives at: 
\begin{align}
\label{eq:alphaphi}
 \sum_{m=0}^\infty \partial_{\varphi} K_{m}^\dagger \partial_{\varphi}
 K_{m}  & 
 =\left[\eta - \alpha(1-\eta) \right]^2  \hat{n}^2  +  \left[ \eta(1-\eta))(1+\alpha)^2 + 2\beta(\alpha(1-\eta) - \eta)\right] \hat{n} + \beta^2, \\
  \sum_{m=0}^\infty \partial_{\eta} K_{m}^\dagger \partial_{\eta}K_{m}
&=   \frac{1}{4 \eta (1-\eta)} \hat{n}.
\end{align}
We see that when we take the expectation values of the above operators with the state $\rho$, as in~\eqref{eq:probeincompboundexp}, we will obtain formulas that can be written in terms of first and second moments of the photon number operator $\langle \hat{n} \rangle$, $\langle \hat{n}^2 \rangle$, or equivalently first moment $\langle \hat{n} \rangle$ and the variance $\Delta^2 {n} = \langle \hat{n}^2 \rangle - \langle \hat{n} \rangle^2$.

To compute a useful bound, we need to minimize the expression in \eqref{eq:probeincompboundexp} over different Kraus representation, which in our case amount to minimization over $\alpha$ and $\beta$. Note that only part involving derivatives over $\varphi$ \eqref{eq:alphaphi} depends on $\alpha$, $\beta$. The dependence is quadratic and hence we can minimize the expectation value of this term over $\alpha$ and $\beta$ explicitly and obtain (see also \cite{Escher2011}):
\begin{equation}
    \min_{\alpha,\beta} \Tr \left[ \rho \sum_{m=0}^\infty \partial_{\varphi} K_{m}^\dagger \partial_{\varphi}
 K_{m}  \right] =  \frac{\eta \langle{ \hat{n}\rangle} }{1 - \eta + \frac{\langle \hat{n} \rangle}{\Delta^2 n} \eta},
\end{equation}
where the optimal choice of parameters corresponds to 
\begin{equation}
\alpha = \frac{\eta(\Delta^2 n - \langle \hat{n} \rangle )}{\Delta^2 n (1-\eta) + \langle \hat{n} \rangle \eta},\quad \beta = \frac{\langle \hat{n} \rangle^2 \eta}{\Delta^2 n (1-\eta)+ \langle \hat{n} \rangle \eta}.
\end{equation}
The probe-incompatibility bound \eqref{eq:probeincompboundexp}, therefore, reads: 
\begin{equation}
    \max_{\rho} \mathcal{F}[\Lambda_{\blambda}(\rho)] \leq \frac{1}{2} \left( \frac{1}{F_{\varphi}^{\mathrm{(max),(1)}}}  \frac{4 \eta \langle{ \hat{ n} \rangle} }{1 - \eta + \frac{\langle \hat{n} \rangle}{\Delta^2 n} \eta} +  \frac{1}{F_\eta^{\mathrm{(max)}}}   \frac{ \langle \hat{n} \rangle}{\eta (1-\eta)}  \right).  
\end{equation}
If we now substitute the formulas for asymptotically saturable single parameter bounds $F_{\varphi}^{\mathrm{(max),(1)}}$ \eqref{upper_phase} and $F_\eta^{\mathrm{(max)}}$\eqref{upper_loss}, we get:  
\begin{equation}
 \max_{\rho} \mathcal{F}[\Lambda_{\blambda}(\rho)] \leq \frac{1}{2} \frac{\langle \hat{n} \rangle}{N} 
 \left( \frac{1}{1 + \frac{\langle \hat{n} \rangle}{\Delta^2 n} \frac{\eta}{1-\eta}} + 1 \right),
\end{equation}
where $N$ is the maximal number of photons allowed.  We see that, in order to have the upper bound equal to 1 (and hence satisfy necessary condition for probe compatibility), we need to satisfy two conditions:
\begin{equation}
\label{eq:neccondprobe}
    \frac{\langle \hat{n} \rangle}{N} \rightarrow 1, \quad \frac{\langle \hat{n} \rangle}{\Delta^2 n} \rightarrow 0.
\end{equation}
The first condition means that we should use states of light that on average contain the maximal amount of photons in the sensing arm that are allowed in the problem, while the second condition requires the photon number statistics in the sensing arm to be super-Poissonian. 

One might wonder if the two conditions are not contradictory.
The example of the state below shows that they are not contradictory: there are single-mode states that satisfy both conditions in the asymptotic limit of large $N$ (this does not mean that this state actually offers the optimal performance, only the existence of states with such property).

Indeed, consider a state:
\begin{equation}
\ket{\psi_N} = \frac{1}{\sqrt{2}}\left( \ket{N}+ \ket{N - N^\alpha} \right),  \quad \frac{1}{2}< \alpha < 1.
\end{equation}
Note that in this case:
\begin{equation}
\langle n \rangle  = N - \frac{1}{2} N^\alpha, \quad \Delta^2 n = \frac{1}{4} N^{2\alpha}.
\end{equation}
Clearly, in the limit $N \rightarrow \infty$ the necessary conditions for probe-compatibilty~\eqref{eq:neccondprobe} are satisfied. 
\subsection{General single and two-mode states}
\label{probe_incomp_nphotons} 
Having found necessary conditions for probe-compatibility, we may now move on to look for the actual states that offer optimal performance in scenarios (1) and (2).  We will perform the maximization of $\mathcal{F} (\rho_{\bm \lambda})$ using the ISS algorithm  to find the optimal states, and to determine the presence or absence of probe incompatibility.
After that, in Sec.~\ref{probe_incomp_gauss}, we will consider classes of single and two-mode Gaussian states, comparing their performance with the results obtained from the ISS maximization.

First, let us define two families of states: a general single-mode state with up to $N$ photons and a general two-mode state with $N$ photons, respectively given by:
\begin{equation}
    | \psi^{(1)}_N \rangle = \sum^N_{n=0} c^{(1)}_n | n \rangle_{a_1} 
    \label{psiN_sm}
\end{equation}
\begin{equation}
    | \psi^{(2)}_N \rangle = \sum^N_{n=0} c^{(2)}_n | n \rangle_{a_1} | N - n \rangle_{a_2} .
    \label{psiN_tm}
\end{equation}
For these classes of states, we assume $\tau_{\mathrm{in}}=1$, since the first state is just a single-mode state and the action of the input beam splitter in the two-mode state may be included in the definition of the state by the coefficients $c^{(2)}_n$.
In other words, these states can be directly chosen as the inputs for schemes (1) and (2) in Fig.~\ref{fig:3schemes}.
Now, the objective is to perform the optimization for these two classes of states by determining the coefficients $c^{(1)}_n$ and $c^{(2)}_n$ that maximize the normalized QFI given by Eq.~\eqref{probe_incomp}, which in this case reads explicitly 
\begin{equation}
    \mathcal{F} (\rho_{\bm \lambda}) = \frac{1}{2} \left( \frac{F_{\varphi \varphi} (\rho_{\bm \lambda})}{F_\varphi^{(\text{max})}} + \frac{F_{\eta \eta} (\rho_{\bm \lambda})}{F_\eta^{(\text{max})}} \right)
    \label{probe_incomp_phase_loss}.
\end{equation}
The maximization of this quantity, which assesses the fundamental probe incompatibility of the channel, is numerically performed with the multiparameter ISS method introduced in Sec.~\ref{preliminaries}.
Here, we give a brief explanation of the form of the states involved, while the implementation of the ISS method for this specific problem is described in detail in~\ref{optimization}.

Let us denote the Fock states as $| n \rangle_{a_k} := (\hat{a}^\dagger_k)^n / \sqrt{n!} |0\rangle$.
To implement the ISS algorithm for the optimization over the states $| \psi^{(1)}_N \rangle$ and $| \psi^{(2)}_N \rangle$, we need to rewrite the expansions given in Eqs.~\eqref{unitary_evol} and \eqref{non_unitary_evol} in the Schrödinger picture, in order to find the Kraus operators such that $\Lambda(| \psi^{(j)} \rangle \langle \psi^{(j)}_N|)=\sum_m K^{(j)}_m | \psi^{(j)} \rangle \langle \psi^{(j)}_N| (K^{(j)}_m)^\dagger$. In this manner, we have~\cite{Dorner2009}:
\begin{equation}
    | n \rangle_{a_1} = \sum^n_{m=0} \sqrt{B^{n}_m} \text{e}^{i n \varphi} | n - m \rangle_{\hat{b}_1} | m \rangle_{e_1} 
    \qquad
    B^{n}_m = \binom{n}{m} \eta^{n-m}(1-\eta)^m ,
    \label{transform_phase_loss}
\end{equation}
where $|m \rangle_e$ is a Fock state for the virtual mode.
Then, replacing Eq.~\eqref{transform_phase_loss} in Eq.~\eqref{psiN_sm} and tracing out over the virtual subspace $| m \rangle_e$, we obtain the single-mode state at the output:
\begin{equation}
    \rho^{(1)}_N = \sum^N_{m=0} | \psi^{(1)}_m \rangle \langle \psi^{(1)}_m | 
    \qquad
    | \psi^{(1)}_m \rangle = \sum^N_{n = m} c^{(1)}_n \sqrt{B^{n}_m} \text{e}^{i n \varphi} | n - m  \rangle_{a_1} | 0 \rangle_{a_2} 
    \label{rho1_output}
\end{equation}
and doing the same with Eq.~\eqref{psiN_tm}, we obtain the two-mode state at the output:
\begin{equation}
    \rho^{(2)}_N = \bigoplus^N_{m=0} | \psi^{(2)}_m \rangle \langle \psi^{(2)}_m |
    \qquad
    | \psi^{(2)}_m \rangle = \sum^N_{n = m} c^{(2)}_n \sqrt{B^{n}_m} \text{e}^{i n \varphi} | n - m  \rangle_{a_1} | N - n \rangle_{a_2} .
    \label{rho2_output}
\end{equation}

\begin{figure}[t]
    \includegraphics[width=0.99\columnwidth]{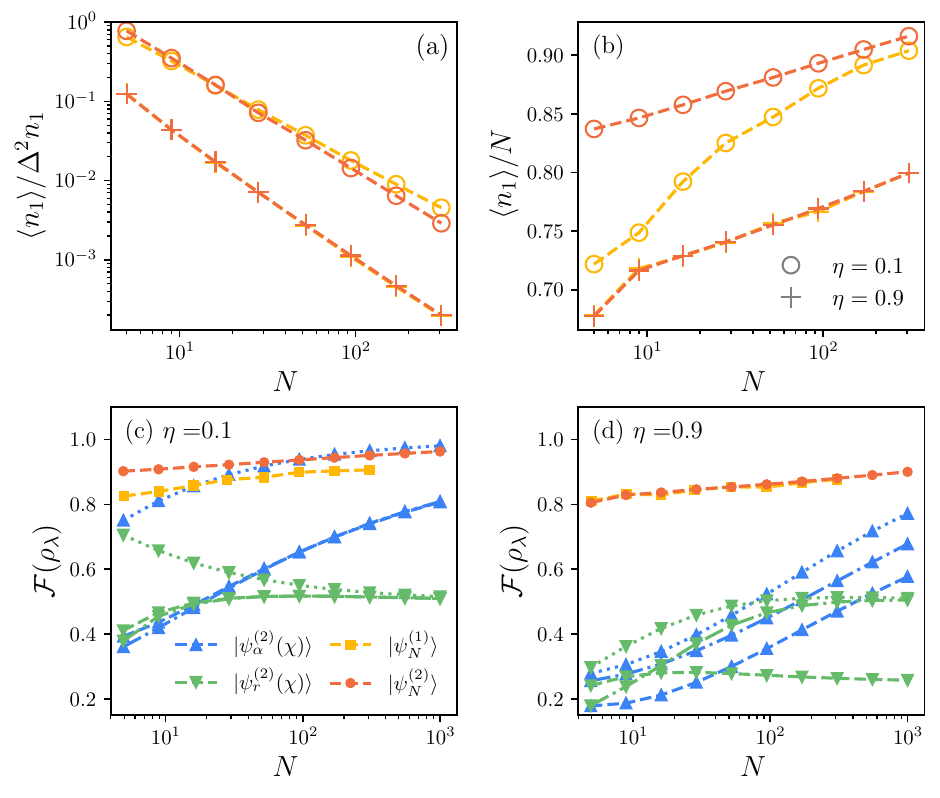}
    \caption{\label{gaussian_vs_optimization}
    Panel (a) shows the photon number variance $\langle n_1 \rangle / \Delta^2 n_1 $; panel (b) the photon number average $\langle n_1 \rangle / N$ of the optimal states $| \psi^{(1)}_N \rangle$ and $| \psi^{(2)}_N \rangle $ obtained from the ISS optimization method.
    Panels (c) and (d) show the normalized QFI $\mathcal{F} (\rho_{\bm \lambda})$ for the optimized states $| \psi^{(1)}_N \rangle$, $| \psi^{(2)}_N \rangle$; the Gaussian states with strong displacement $| \psi^{(2)}_\alpha (\chi) \rangle$ and strong squeezing $| \psi^{(2)}_r (\chi) \rangle$. For the Gaussian states we have $\chi=0$ (dashed), $\chi=\pi/4$ (dash-dotted) and $\chi=\pi/2$ (dotted).}
\end{figure}

In Fig. \ref{gaussian_vs_optimization} we show the normalized QFI for both states, $\mathcal{F}(\rho^{(1)}_N)$ and $\mathcal{F}(\rho^{(2)}_N)$, resulting from the ISS optimization.
Additionally, the off-diagonal elements of the QFI matrix vanish for the optimal two-mode state and become negligible as $N$ increases for the optimal single-mode state.
We remark that~\cite{Crowley2014} considered a gradient method to optimize over two-mode states of the form in Eq.~\eqref{psiN_tm}, minimizing the combined variances for photon numbers up to only $N=200$.
In contrast, our iterative see-saw method is able to achieve this optimization problem for photon numbers up to $N=1000$, as shown in Fig.~\ref{gaussian_vs_optimization}.

\begin{figure}[h]
\includegraphics[width=0.9 \columnwidth]{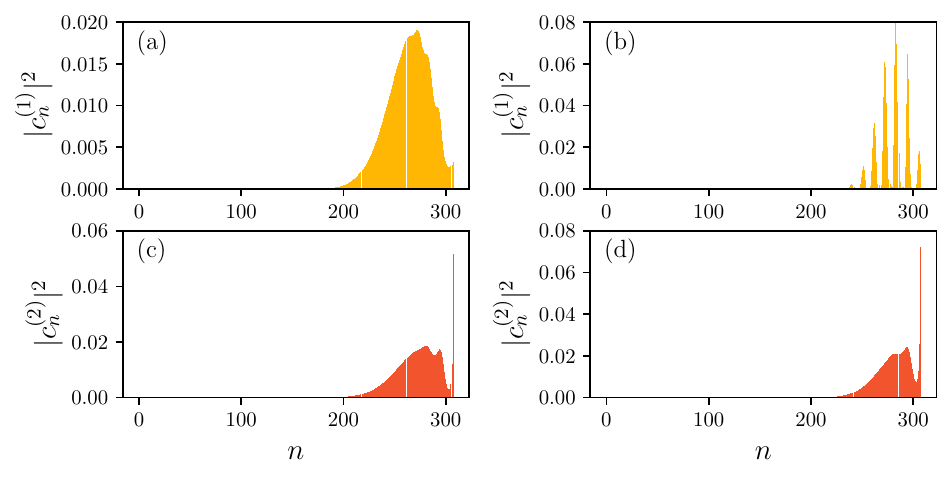}
\caption{\label{histogram_optimalstate}
    The top panels show the photon-number distribution of the optimal single-mode state $| \psi^{(1)}_{307} \rangle$: (a) for phase estimation only, and (b) for simultaneous phase and loss estimation.
    The bottom panels show the photon-number distribution of the optimal two-mode state $| \psi^{(2)}_{307} \rangle$: (c) for phase estimation only, and (d) for simultaneous phase and loss estimation.
    In all the graphs we fix $\eta=0.1$.}
\end{figure}

We can further understand how these states overcome probe incompatibility by analyzing the photon number variance, also shown in Fig.~\ref{gaussian_vs_optimization}.
The intuition that the photon-number variance must be sufficiently large to enable precise optimal phase estimation stems from noiseless phase estimation, but we have also demonstrated rigorously in Sec.~\ref{sec:necprobecomp} that this condition holds even for lossy phase estimation.
Indeed, from Fig.~\ref{gaussian_vs_optimization}(a), we conclude that both states exhibit a large photon number variance, as $\langle \Delta^2 n_1 \rangle / N$ follows a super-Poissonian scaling, which benefits phase estimation.

Furthermore, as a necessary condition for probe-compatibility we also require the states to have mean photon number in the sensing arm approaching the maximal number.
Again, Fig.~\ref{gaussian_vs_optimization}(b) shows that the average photon number approaches that of a Fock state, since $\langle \hat{n}_1 \rangle$ asymptotically approaches $N$.

However, the optimal single-mode state for loss estimation is a Fock state~\cite{Adesso2009}, which inherently has zero photon number variance.
We, therefore, may expect to have some non-trivial trade-offs here.
Interestingly, these trade-offs seem to affect only the single-mode scenario (1).
Intuitively, the presence of the reference mode allows one to have the best of both worlds---large variance of photon number in the sensing mode, but at the same time precise information on the number of photons entering this mode, thanks to photon-number entanglement between sensing and reference modes.

This observation is reflected in the fact that the optimal states in scenarios (1) and (2) exhibit different sensing-mode photon number distributions.
This is highlighted in Fig.~\ref{histogram_optimalstate}, where we show the histogram of the photon number distribution of optimal states for both simultaneous estimation and phase estimation, choosing the exemplary value $N=307$.
In the single-mode case (1), the optimal state for simultaneous estimation exhibits a teeth-like structure, which is necessary for accurately estimating loss, as shown in Fig.~\ref{histogram_optimalstate}(b)---the intuition being that the separation of the teeth allows one to better determine the original number of photons based on the number of photons detected.
As a result, despite the larger photon-number variance, we obtain better loss sensitivity.
In fact, this structure is not required in the two-mode case (2), due to the presence of the additional reference beam.

\me{To conclude this discussion, we mention that the ISS method for the maximization of Eq.~\eqref{probe_incomp} can be complemented and validated by computing the upper bound in Eq.~\eqref{eq:probeincompbound}.
This is numerically efficient and amounts to solving an SDP~\cite{Albarelli2022}.
As mentioned in Sec.~\ref{sec:probeincompbound}, the bound in Eq.~\eqref{probe_incomp} is saturable by probe states that are entangled with noiseless ancillas.
Thus, the bound will not be tight for the single mode scenario, scheme (1) in Fig.~\ref{fig:3schemes}, where no ancillary noiseless mode is available.
On the other hand, it is an alternative way to compute the probe incompatibility quantifier for scheme (2).
Indeed, we have checked the results of these different numerical methods match.
In appendix~\ref{app:KrausMin} we provide more details on this algorithm, which allows to reach larger values of $N$, since restricting to diagonal operators in the Fock basis suffices for its evaluation, analogously to single-parameter lossy phase estimation~\cite[App.~G]{Zhou2020}.}

\subsection{\label{probe_incomp_gauss} Gaussian states}

Let ${\bm A} = (\hat{a}_1, \hat{a}_2, \hat{a}^\dagger_1, \hat{a}^\dagger_2)$ be the coordinate vector of a two-mode Gaussian state, that satisfies the commutation relation $\left[\hat{A}_i, \hat{A}^\dagger_j \right] = i \Omega_{i j}$, where $\Omega$ is called the symplectic matrix, defined by:
\begin{equation}
    \Omega = 
    \begin{pmatrix}
    I_2 & 0_2 \\   
    0_2& - I_2
    \end{pmatrix},
\end{equation}
where $I_2$ is the $2{\times}2$ identity matrix.
Any Gaussian state is fully described by its displacement vector ${\bm d}$ and covariance matrix $\sigma$ defined respectively as follows~\cite{Adesso2014}:
\begin{equation}
    \sigma_{ij} = \langle \{ \hat{A}_i , \hat{A}^\dagger_j \} \rangle - 2 \langle  \hat{A}_i \rangle \langle \hat{A}^\dagger_j \rangle
    \qquad
    d_i = \langle \hat{A}_i  \rangle 
    \label{cov_disp_in}
\end{equation}
where the covariance matrix satisfies the condition $\sigma + \Omega  \geq  0$.
The evolution of a Gaussian state $| \psi \rangle$ into the output state $\rho_{\bm \lambda} = \Lambda_{\bm \lambda}(| \psi \rangle \langle \psi | )$ is mapped by transforming the covariance matrix and displacement vector as: 
\begin{equation} 
    \sigma_{\bm \lambda} = \sqrt{\bm \eta} \left( \mathcal{U} \sigma  \mathcal{U}^\dagger - I_4 \right) \sqrt{\bm \eta} + I_4 
    \qquad
    {\bf d}_{\bm \lambda} = \sqrt{\bm \eta} \hspace{0.5mm} \mathcal{U} {\bf d} ,
    \label{cov_disp_out}
\end{equation}
where the $\mathcal{U}$ gives the unitary evolution due to the action of the input beamsplitter and the phase shift, being  constructed from the $2$-dimensional matrices $U$ in Eq.~\eqref{unitary_evol} as follows:
\begin{equation}
    \mathcal{U} = 
    \begin{pmatrix}
    U & 0_2 \\
    0_2 & U^*
    \end{pmatrix},
\end{equation}
and the non-unitary evolution due to the loss is recast as the matrix $\sqrt{\bm \eta} = \text{diag}(\sqrt{\eta}, 1, \sqrt{\eta}, 1)$, obtained from Eq.~\eqref{non_unitary_evol}.
Eq.~\eqref{cov_disp_out} is proven by the direct application of Eqs.~\eqref{unitary_evol}, \eqref{non_unitary_evol} in the definition of the covariance matrix and displacement vector, given by Eq.~\eqref{cov_disp_in}.
Following~\cite{Safranek2019} (see also the previous works~\cite{Monras2013,Gao2014a,Serafini2023}), the QFI matrix of a Gaussian state can be calculated in terms of the covariance matrix and displacement vector at the output.

Additionally, when dealing with states with indefinite photon number, the ultimate quantum precision limit for the phase estimation, given in Eq.~\eqref{upper_phase}, should be interpreted with some caution.
When $N$ is replaced by the average photon number $ \langle N \rangle $, this upper bound holds only when no additional reference beam is considered.
This is because the QFI is obtained from the optimization over all possible POVMs, which also include detection schemes that uses additional reference beams (e.g., homodyne and heterodyne detections), without accounting the energy spent in the reference beam~\cite{Jarzyna2012}.
 
\subsubsection{\label{section_gauss_sm} Single-mode Gaussian states.}

We start from the single-mode case, in which the photons are injected only in the probe mode and we do not have access to the ancillary mode.
The single-mode squeezing operator is defined as:
\begin{equation}
    S^{(1)} (r) =   \text{exp}\left( \frac{r}{2} \text{e}^{-i \theta} \hat{a}^2_1 - \frac{r}{2} \text{e}^{i \theta}  (\hat{a}^\dagger_1)^2 \right) ,
    \label{squeezing}
\end{equation}
where $\bar{N}_r = \hspace{1mm} \sinh^2 r$ gives the average photon number generated and $\theta$ is the direction of the squeezing in the phase space.
For a single mode, the most general pure state is a displaced squeezed state,
generated from the action of the single-mode squeezing operator in the vacuum, followed by the action of the displacement operator defined in Eq.~\eqref{displacement} in the first mode, as follows:
\begin{equation}
    | \psi^{(1)}_G \rangle =D^{(1)}(\alpha) S^{(1)} (r) | 0, 0 \rangle ,
    \label{gaussian_singlemode}
\end{equation}
with average total photon number given  by $\bar{N} = \bar{N}_r + \bar{N}_\alpha$.
Thus, in the coordinates ${\bm A} = (\hat{a}_1, \hat{a}_2, \hat{a}^\dagger_1, \hat{a}^\dagger_2)$, the probe state given in Eq.~\eqref{gaussian_singlemode} has the covariance matrix:
\begin{equation}
    \sigma^{(1)}_G = 
    \begin{pmatrix}
    \cosh (2r)  & 0 & -\sinh (2r) \hspace{0.5mm} \text{e}^{i \theta_1}  & 0 \\
    0 & 1 & 0 & 0 \\
    -\sinh (2r) \hspace{0.5mm} \text{e}^{-i \theta_1} & 0 & \cosh (2r) & 0 \\
    0 & 0 & 0 & 1
    \end{pmatrix} ,
    \label{cov_singlemode}
\end{equation}
and the displacement vector:
\begin{equation}
    {\bf d}^{(1)}_G =
    \begin{pmatrix}
    \alpha \hspace{0.5mm} \text{e}^{i \mu} &
    0 &
    \alpha \hspace{0.5mm} \text{e}^{-i \mu}  &
     0
    \end{pmatrix}^T .
    \label{displacement_vector}
\end{equation}

Thus, the covariance matrices and displacement vectors at the output are found by replacing Eqs.~\eqref{cov_singlemode},~\eqref{displacement_vector} in Eq.~\eqref{cov_disp_out} with $\tau_{\mathrm{in}} = 1$, since we are now working only with the probe mode.
To simplify the analysis, here we focus on the asymptotic limits of $F_{\varphi \varphi} (\rho^{(1)}_G)$ and $F_{\eta \eta}(\rho^{(1)}_G)$, as their full expressions are quite involved (see \ref{gaussian_calculations} for reference).
In order to optimize the ratio of energy devoted to displacement and squeezing, we introduce the coefficient $0 < p < 1$, which governs the energy contributions.
We begin by considering the state with strong displacement, denoted by $| \psi^{(1)}_\alpha \rangle$, defined as the state in Eq.~\eqref{gaussian_singlemode} with $\bar{N}_\alpha = \bar{N} - \bar{N}^p$ and $\bar{N}_r = \bar{N}^p$.
For this state, we have:
\begin{equation}
    \frac{F_{\varphi \varphi} (\rho^{(1)}_\alpha)}{F^{\text{(max)}}_\varphi} \overset{N \rightarrow \infty}{\longrightarrow} \sin^2\left(\frac{\theta_1-2\mu}{2} \right)
    \qquad
    \frac{F_{\eta \eta} (\rho^{(1)}_\alpha)}{F^{\text{(max)}}_\eta} \overset{N \rightarrow \infty}{\longrightarrow} 
    \cos^2\left(\frac{\theta_1-2\mu}{2} \right) .
    \label{qfi_sm_alpha}
\end{equation}
Therefore, the incompatibility of using a single-mode Gaussian state for the simultaneous estimation of phase and loss becomes evident.
Achieving the ultimate quantum precision bound for phase estimation requires the state being phase squeezed (i.e., $\theta_1 - 2 \mu = \pi$), which suppresses the loss estimation. In contrast, achieving the ultimate quantum precision bound for loss estimation requires the state being amplitude squeezed (i.e., $\theta_1 - 2 \mu = 0$), which suppresses the phase estimation.
In~\cite{Monras2007} the optimal phase choice for the loss estimation agrees with the Eq.~\eqref{qfi_sm_alpha}, however there was no mention about the fundamental bound~\eqref{upper_loss}, since it was not known at the time. Following this, we denote the state with strong squeezing as $| \psi^{(1)}_r \rangle$, that is defined as the state in Eq.~\eqref{gaussian_singlemode} with $\bar{N}_r = \bar{N} - \bar{N}^p$ and $\bar{N}_\alpha = \bar{N}^p$. For this state, we have:
\begin{equation}
    \frac{F_{\varphi \varphi} (\rho^{(1)}_r)}{F^{\text{(max)}}_\varphi} \overset{N \rightarrow \infty}{\longrightarrow} 1
    \qquad
    \frac{F_{\eta \eta} (\rho^{(1)}_r)}{F^{\text{(max)}}_\eta} \overset{N \rightarrow \infty}{\longrightarrow} 0 ,
    \label{qfi_sm_r}
\end{equation}
which achieves the ultimate quantum precision bound for the phase estimation but fails to achieve even the SQL scaling for the loss estimation.

In addition, the correlation term of the QFI matrix is given by:
\begin{equation}
    F_{\varphi \eta} \big( \rho^{(1)}_G \big) =
    \frac{4  \eta \sin (\theta_1 -2 \mu )  \sqrt{\bar{N}_r (\bar{N}_r+1)} \bar{N}_\alpha}{4 (\eta -1) \eta  \bar{N}_r-1},
    \label{qfi_sm_correlation}
\end{equation}
which is zero in the two extremal cases $\theta_1 - 2 \mu = 0$ (squeezing and displacement in the same direction) or $\theta_1 - 2 \mu = \pi$ (squeezing and displacement in the opposite direction).
In the strong displacement regime, each case corresponds to the optimal value for one of the QFIs, while the other QFI vanishes.
Furthermore, the maximum value of the correlation is attained when $\theta_1 - 2 \mu = \pi/2$, which corresponds to the case when both parameters are estimated with the same importance, resulting in $F_{\varphi \varphi} \big( \rho^{(1)}_G \big) = F_{\eta \eta} \big( \rho^{(1)}_G \big)$. Finally, we observe that this correlation becomes weaker in the case of large losses ($\eta \approx 0$).
We mention that~\cite{Pinel2013} also discusses parameter estimation using single-mode Gaussian states, including phase and loss, but even there the results were not compared to the  ultimate quantum limits.
Indeed, Eqs.~\eqref{qfi_sm_alpha} and \eqref{qfi_sm_r} can also be derived from Eqs.~(16) and (22) in~\cite{Pinel2013}, which however lacks any discussion about this type of energy distribution, the fundamental bounds and these fundamental tradeoff.

As previously discussed, some intuition about probe incompatibility can be gained  by examining the photon number variance of the probe mode.
For the state considered here, defined in Eq.~\eqref{gaussian_singlemode}, we have:
\begin{equation}
    \Delta^2 n_1 = 2 \bar{N}_r (\bar{N}_r + \bar{N}_\alpha + 1) + \bar{N}_\alpha - 2 \bar{N}_\alpha \sqrt{\bar{N}_r (\bar{N}_r + 1)} \cos \left( \theta - 2 \mu \right).
\end{equation}
According to Eq.~\eqref{qfi_sm_alpha}, in the strong displacement regime the optimal phase relation for phase estimation is $\theta - 2\mu = \pi$, which maximizes the photon number variance.
In contrast, the optimal phase relation for the loss estimation is $\theta - 2\mu = 0$, which minimizes the photon number variance. Note that, in the last case the photon number variance has a scaling $\Delta^2 n_1  \sim \bar{N}^{p/2}$, which cannot satisfy the necessary condition for probe compatibility, given by Eq.~(\ref{eq:neccondprobe}).
Additionally, the single-mode Gaussian state cannot reproduce the teeth-like structure of the optimal single mode state $| \psi^{(1)}_N \rangle$, which is needed for estimating loss in the simultaneous estimation scenario, as shown in Fig.~\ref{histogram_optimalstate}(b).

\subsubsection{\label{tm} Two-mode Gaussian state.}
As demonstrated in the previous section, a single-mode state exhibits probe  incompatibility.
To overcome this limitation, it is necessary to make use of the ancillary mode and incorporate some degree of entanglement into the state.
First, let us define the generalized two-mode squeezing operator as follows:
\begin{equation}
    S^{(2)}_\chi(r) 
     = \text{exp} \Big[
    \frac{r}{2} \hspace{1mm}  \text{cos} \chi \hspace{1mm} \left(
    \text{e}^{-i \theta_1 } \hat{a}^2_1 +
    \text{e}^{-i \theta_2 } \hat{a}^2_2
    \right) 
     + r \hspace{1mm} \text{sin} \chi \hspace{1mm} \text{e}^{-i \theta} \hat{a}_1 \hat{a}_2 - h.c. \Big],
     \label{general_squeez}
\end{equation}
where here $\bar{N}_r = 2 \hspace{1mm} \sinh^2 r$  gives the average total photon number generated  from the squeezing in both modes. The phase $\chi$ plays an important role in the preparation of the input state, as it determines when this operator generates two single-mode squeezed states ($\chi = 0$), a two-mode squeezed state ($\chi = \pi/2$), or a state between these two ($0 < \chi < \pi/2$). Indeed, this kind of operator can be generated from the interference of two single-mode squeezed states in a beam splitter \cite{Yeoman1993}, where the phase $\chi$ is controlled by changing the transmissivity and reflectance phase of this beamsplitter. Finally, our probe state is:
\begin{equation}
    | \psi^{(2)}_G (\chi) \rangle = D^{(1)}(\alpha) S^{(2)}_\chi(r) | 0, 0 \rangle ,
    \label{gaussian_twomode}
\end{equation}
with average total photon number given  by $\bar{N} = \bar{N}_r + \bar{N}_\alpha$ and with displacement also acting only in the first mode, as defined in Eq.~\eqref{displacement}.
In this way, in the coordinates ${\bm A} = (\hat{a}_1, \hat{a}_2, \hat{a}^\dagger_1, \hat{a}^\dagger_2)$, the probe state given in Eq.~\eqref{gaussian_twomode} has the covariance matrix:
\begin{equation}
    \sigma^{(2)}_G = 
    \begin{pmatrix}
    \cosh (2r) \hspace{0.5mm} I_2  & 
    -\sinh (2r) \hspace{0.5mm} R_\chi \\
    -\sinh (2r) \hspace{0.5mm} R^*_\chi  & 
    \cosh (2r) \hspace{0.5mm} I_2
    \end{pmatrix} 
    \qquad
    R_\chi = 
    \begin{pmatrix}
    \cos \chi \hspace{0.5mm} \text{e}^{i \theta_1}  & 
    \sin \chi \hspace{0.5mm} \text{e}^{i \theta} \\
    \sin \chi \hspace{0.5mm} \text{e}^{i \theta} & 
    \cos \chi \hspace{0.5mm} \text{e}^{i \theta_2}
    \end{pmatrix} ,
    \label{cov_twomode}
\end{equation}
and the displacement vector is the same as for the single-mode state, ${\bf d}^{(2)}_G = {\bf d}^{(1)}_G$, since we considered displacement only in the probe mode.

Thus, the corresponding covariance matrices and displacement vectors at the output are found by replacing Eqs.~\eqref{cov_twomode}, \eqref{displacement} in Eq.~\eqref{cov_disp_out}.
Analogously to the single-mode Gaussian state, we introduce the coefficient $0 < p < 1$  that defines the energy contributions.
First, we denote the two-mode state with strong displacement by $| \psi^{(2)}_\alpha (\chi) \rangle $, as the state in Eq.~\eqref{gaussian_twomode} with $\bar{N}_\alpha = \bar{N} - \bar{N}^p$ and $\bar{N}_r = \bar{N}^p$.
This class of states can asymptotically attain the ultimate precision limit for both phase and loss estimation as $\tau_{\mathrm{in}} \rightarrow 1$.
However, this approach must be implemented with caution, as discussed in the following sections.

First, let us consider the probe state $| \psi^{(2)}_\alpha (0) \rangle$, which consists of two single-mode squeezed states with strong displacement. In that case, we cannot simply assume that $\tau_{\mathrm{in}} = 1$, as this returns to the single-mode case described in section \ref{section_gauss_sm}. Then, in order to have access to the probe mode, we consider a transmissivity that goes asymptotically to one, in the form $\tau_{\mathrm{in}} = 1 - 1/\bar{N}^q$, with $0<q<1$. In that way, we have the following limits for the phase QFI:
\begin{equation}
    \frac{F_{\varphi \varphi} ( \rho^{(2)}_\alpha (0) )}{F^{\text{(max)}}_\varphi} \overset{N \rightarrow \infty}{\longrightarrow} 
    \begin{cases} 
    1 & \text{for } q < p, \\
        1-\frac{\eta  \cos ^2\left(\frac{\theta_1-2\mu}{2}\right)}{1 + (1-\eta ) \cos (\theta_1-\theta_2)} & \text{for } q = p, \\
    \sin^2\left(\frac{\theta_1-2\mu}{2} \right) & \text{for } q > p,
    \end{cases}
    \label{qfi_phase_sm}
\end{equation}
and for the loss QFI:
\begin{equation}
    \frac{F_{\eta \eta} ( \rho^{(2)}_\alpha (0) )}{F^{\text{(max)}}_\eta} \overset{N \rightarrow \infty}{\longrightarrow} 
    \begin{cases} 
    1 & \text{for } q < p, \\
    1-\frac{\eta  \sin ^2\left(\frac{\theta1-2 \mu}{2})\right)}{1 + (1-\eta ) \cos (\theta1-\theta_2)} & \text{for } q = p, \\
    \cos^2\left(\frac{\theta_1-2\mu}{2} \right) & \text{for } q > p,
    \end{cases}
    \label{qfi_loss_sm}
\end{equation}
From the last two equations, we conclude that the ultimate quantum precision bounds for phase and loss are achieved simultaneously when $q<p$, which means that the transmissivity is converging sufficiently slowly to one, ensuring that the squeezing contributions from the two beams always interfere, i.e., $\lim_{\bar{N} \rightarrow \infty} (1-\tau_{\mathrm{in}})  \bar{N}_r = \infty$. In contrast, when $q>p$ the transmissivity is converging sufficiently fast to one in such a way that the squeezing contributions from the two beams do not interfere asymptotically, i.e., $\lim_{\bar{N} \rightarrow \infty} (1-\tau_{\mathrm{in}})  \bar{N}_r = 0$.
This last case reproduces the results of the single-mode state given by Eq.~\eqref{qfi_sm_alpha} and then the same incompatibility problem appears.
Finally, when $p=q$ the ultimate precision limit of both parameters can be simultaneously achieved  in the regime of large losses ($\eta \approx 0$).

In Fig.~\ref{gaussian_vs_optimization} we show the normalized QFI for states of the form $\mathcal{F}(\rho^{(2)}_\alpha (0))$, considering two values of loss ($\eta = 0.9, 0.1$), and comparing the results with the non-Gaussian optimal states obtained numerically from the ISS method. Finally, in the strong displacement regime, the off-diagonal QFI matrix element is given by:
\begin{eqnarray}
    F_{\varphi \eta} \big( \rho^{(2)}_\alpha (0) \big)  
    &\approx&
    \frac{4 \eta \tau^2_{in} \sin(\theta_1-2 \mu)  \sqrt{\bar{N}_r(\bar{N}_r + 1)} \bar{N}_\alpha}{\text{den}} +
    \label{qfi_correlation_sm}\\
    && - \frac{4 \eta \tau_{\mathrm{in}} (1 - \tau_{\mathrm{in}}) \sin(\theta_2-2 \mu) \sqrt{\bar{N}_r(\bar{N}_r + 1)} \bar{N}_\alpha}{\text{den}},
    \nonumber
\end{eqnarray}
where we retain only the displacement contribution of the QFI matrix, see for instance Eq.~\eqref{qfi_general_gaussian}; and the denominator is given by: $\text{den} = 4 \eta  (1-\eta ) \bar{N}_r+8 (1-\eta)^2 (1-\tau_{\mathrm{in}}) \tau_{\mathrm{in}} \bar{N}_r (\bar{N}_r+1) \cos (\theta_1-\theta_2)+8 (1-\eta)^2 (1-\tau_{\mathrm{in}}) \tau_{\mathrm{in}} \bar{N}_r(\bar{N}_r + 1)+1 $.
First, setting the input transmissivity in the form  $\tau_{\mathrm{in}} = 1 - 1/\bar{N}^q$, we recover the same expression for the off-diagonal QFI element of the single-mode state, given in Eq.~(\ref{qfi_sm_correlation}), which vanishes when $\theta_1 - 2 \mu = 0,\pi$.
However, if both squeezing phases are such that $\theta_1 - 2 \mu = \theta_2 - 2 \mu = 0,\pi$, the off-diagonal element of the QFI vanishes for any value of input transmissivity $\tau_{\mathrm{in}}$.

Following this discussion, let us consider the probe state $| \psi^{(2)}_\alpha (\pi/2) \rangle$, which consists of a two-mode squeezed state with strong displacement.
In that case, we can set $\tau_{\mathrm{in}}=1$ due to the intrinsic entanglement already existing in the two-mode state. Then, we have asymptotically:
\begin{equation}
    \lim_{\bar{N} \rightarrow \infty} \frac{F_{\varphi \varphi} ( \rho^{(2)}_\alpha (\pi/2) )}{F^{\text{(max)}}_\varphi} \overset{N \rightarrow \infty}{\longrightarrow} 1
    \qquad
    \lim_{\bar{N} \rightarrow \infty} \frac{F_{\eta \eta} ( \rho^{(2)}_\alpha (\pi/2) )}{F^{\text{(max)}}_\eta} \overset{N \rightarrow \infty}{\longrightarrow} 1
    \label{qfi_phase_loss_tm}
\end{equation}
The corresponding normalized QFI of this state, $\mathcal{F}(\rho^{(2)}_\alpha (\pi/2))$, is shown in Fig.~\ref{gaussian_vs_optimization}, where we conclude that this state (i.e., with $\chi=\pi/2$) performs better than the previously considered one (i.e., with $\chi=0$).

We mention that after an appropriate reparametrization, the previous equations can be derived from the results of previous works~\cite{Dowran2021,Woodworth2022}, which considered a bright two-mode squeezed state (displacement before squeezing) as the probe state to estimate loss. Finally, in the strong displacement regime, the off-diagonal QFI matrix element is given by:
\begin{equation}
    F_{\varphi \eta} \big( \rho^{(2)}_\alpha (\pi/2) \big)  
    \approx
    \frac{8 \cos (\theta -2 \mu ) \eta \tau_{\mathrm{in}} \sqrt{(1-\tau_{\mathrm{in}}) \tau_{\mathrm{in}}} \sqrt{\bar{N}_r(\bar{N}_r+1)} \bar{N}_\alpha}{4 (1-\eta) \Big[ 4 (1-\eta) (1-\tau_{\mathrm{in}}) \tau_{\mathrm{in}}+(\eta -1) (1-2 \tau_{\mathrm{in}})^2 \bar{N}_r -1 \Big] \bar{N}_r -1
    }
    \label{qfi_correlation_tm}
\end{equation}
which is zero when the squeezing and displacement are performed in an orthogonal direction, i.e. $\theta - 2 \mu = \pm \pi/2$.
In addition, in Fig. \ref{gaussian_vs_optimization}, we show the normalized QFI for an intermediate probe state, $| \psi^{(2)}_\alpha (\pi/4) \rangle$  with $\tau_{\mathrm{in}} = 1$, whose performance is between the two previous ones.

Lastly, we consider the probe states with strong squeezing, $| \psi^{(2)}_r (\chi) \rangle $, defined as the state in Eq. (\ref{gaussian_singlemode}) with $\bar{N}_r = \bar{N} - \bar{N}^p$ and $\bar{N}_\alpha = \bar{N}^p$ . The normalized QFI for these states is presented in Fig. \ref{gaussian_vs_optimization} for $\chi=0,\pi/4,\pi/2$. All cases converge to half of the ultimate quantum precision, which can be attributed to preparing the states with half photons in the probe mode and half in the ancillary mode.
From this point onward, we will focus exclusively on the states with strong displacement, as they can reach the ultimate quantum limit for both parameters.

\section{\label{sec_measurement_incomp} Measurement incompatibility for single-mode loss}

\subsection{\me{Fundamental measurement incompatibility of optimal states}}

In the previous section, we identified some classes of states, both Gaussian and non-Gaussian that simultaneously achieve the ultimate quantum precision limits for the estimation of phase and loss.
In this section, we aim to show that measurement incompatibility persists for these states.

To begin, we recall the definition of the expectation value of the average of the commutators of the SLDs, given by Eq.~\eqref{incomp}.
When it is not zero, it implies trade-offs between the precision of different parameters, thus limiting the simultaneous estimation performance when using one single measurement strategy for both parameters.
First, for the probe state $| \psi^{(2)}_N \rangle$ given by Eq.~\eqref{rho2_output} it was proven in~\cite{Crowley2014} that the expectation value of the commutators of the SLDs for phase and loss is given by:
\begin{equation}
     I_{\varphi \eta} ( \rho^{(2)}_N ) =  \frac{i F_{\varphi \varphi} (\rho^{(2)}_N )}{2 \eta} .
     \label{incomp_psin}
\end{equation}
It means that the measurement incompatibility can be surpassed only at the cost of vanishing the phase estimation precision.
Following~\cite{Safranek2019,Nichols2018}, the expectation value $I_{\varphi \eta}$ of any Gaussian state can be calculated in terms of the covariance matrix $\sigma_{\bm \lambda}$ and displacement vector ${\bf d}_{\bm \lambda}$.
Considering the states with strong displacement, we found asymptotically a similar proportionality  relation between $I_{\varphi \eta} ( \rho^{(2)}_\alpha (\chi) )$ and the phase QFI,  fixing the values $\chi=0$ and $\tau_{\mathrm{in}} = 1-1/\bar{N}^q$ we obtain:
\begin{equation}
    \frac{I_{\varphi \eta} ( \rho^{(2)}_\alpha (0) )}{F^{\text{(max)}}_\varphi} \overset{N \rightarrow \infty}{\longrightarrow} 
    \begin{cases} 
     \frac{i}{\eta} & \text{for } q < p, \\
     \frac{i \eta \big[ (1-\eta ) \cos (\theta_1-\theta_2)+1 \big]}{(1-\eta ) \big[ \cos (\theta_1-\theta_2)+1 \big] } & \text{for } q = p, \\
     0 & \text{for } q > p,
    \end{cases}
    \label{incomp_sm}
\end{equation}
and for $\chi=\pi/2$ and $\tau_{\mathrm{in}} = 1$:
\begin{equation}
    \frac{I_{\varphi \eta} ( \rho^{(2)}_\alpha (\pi/2) )}{F^{\text{(max)}}_\varphi} \overset{N \rightarrow \infty}{\longrightarrow} \frac{i}{\eta} .
    \label{incomp_tm}
\end{equation}

Therefore, by examining the fundamental measurement incompatibility given by $I_{\varphi \eta} ( \rho^{(2)}_N )$, we conclude that our optimal states cannot overcome measurement incompatibility.
This highlights the importance of considering the measurement incompatibility indicators $\rh$ and $\roh$ introduced in Eq.~\eqref{r_h}.
These quantities, which are based on the HCRB, establish fundamental bounds that depend only on the probe state.
Additionally, in order to get regularized bounds, we consider the weight matrix $W=\text{diag}(F_\varphi^{(\text{max})}, F_\eta^{(\text{max})})$.

\begin{figure}[t]
\includegraphics[width=1 \columnwidth]{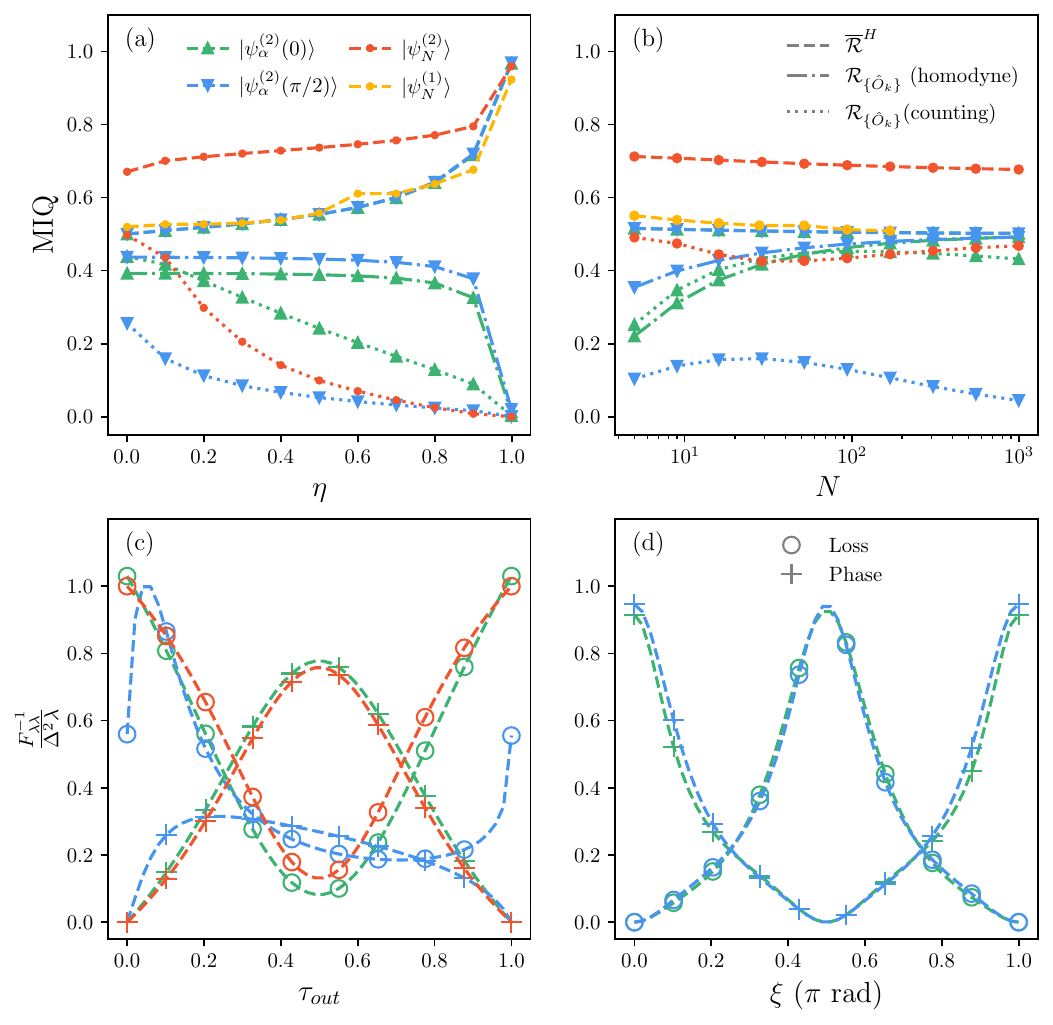}
\caption{
Panels (a) and (b) at the top show the measurement incompatibility quantifiers (MIQ), where we consider the fundamental bound $\roh$ (dashed), the bounds 
$\mathcal{R}_{\{ O_k \}}(\rhol)$ for the photon counting (dotted) and homodyne detection (dash-dot). In panel (a) these bounds are plotted in function of $\eta$ with fixed photon number $N=20$ and in panel (b) they are plotted as a function of $N$ with fixed $\eta=0.1$. For photon counting, we consider the half photon strategy with $\tau_{\mathrm{out}}=1/2$ for the phase estimation and $\tau_{\mathrm{out}}=1$ for the loss estimation. For homodyne detection, we consider the simultaneous estimation strategy with 
$\xi=\pi/4$ and $\tau_{\mathrm{out}}=1$ for both parameters.
In the bottom panels (c) and (d), we show the measurement incompatibility for the independent estimation with the figure of merit being $1-F^{-1}_{\varphi \varphi} / \Delta^2 \varphi$ for the phase estimation (crosses) and $1-F^{-1}_{\eta \eta} / \Delta^2 \eta$ for the loss estimation (crosses), considering photon counting in panel (c) and homodyne detection in panel (d).
In both graphs (c) and (d) we have $N=94$ and $\eta=0.1$.
In all the fours plots, we consider the weight matrix $W=\text{diag}(F^{\text{(max)}}_\varphi, F^{\text{(max)}}_\eta)$ and the energy distributions $\bar{N}_\alpha=\bar{N} - \sqrt{\bar{N}}$, 
$\bar{N}_r = \sqrt{\bar{N}}$ and $\mu=0$ for Gaussian states. 
\label{variances_vs_bound}
}
\end{figure}

In Fig.~\ref{variances_vs_bound}, panels (c) and (d), we plot the measurement incompatibility indicator $\rh$ for the optimal states able to overcome probe incompatibility identified in the previous section.

For Gaussian states we were also able to evaluate numerically the bound $C^H (\rho_{\bm \lambda})$, leveraging the results of~\cite{Chang2025}.
Specifically, in \ref{gaussian_calculations} we show that for large losses ($\eta \approx 0$) the HCRB converges to the upper bound $\overline{C}^H (\rho_{\bm \lambda})$.

Considering the optimized two-mode state with $N$ photons $| \psi^{(2)}_N \rangle$,
when $\eta \neq 1$, from Eq.~\eqref{incomp_psin} we obtain the following limit:
\begin{equation}
    \roh \overset{N \rightarrow \infty}{\longrightarrow} \frac{2}{3},
\end{equation}
and in the same way, considering the two-mode Gaussian states with strong displacement, from Eq.~\eqref{incomp_sm} with $q<p$ and Eq.~\eqref{incomp_tm} we obtain the following limit:
\begin{equation}
    \roh \overset{N \rightarrow \infty}{\longrightarrow} \frac{1}{2} .
\end{equation}
Finally, when considering the optimized single-mode state with cutoff at $N$ photons $| \psi^{(1)}_N \rangle$, we find numerical evidence that it achieves the same measurement incompatibility bound as the Gaussian states considered, as shown in Fig.~\ref{variances_vs_bound}, panels (a) and (b).

\subsection{\me{Measurement incompatibility of practical optical detection schemes}}

To finish our discussion about measurement incompatibility, we  explicitly analyze some paradigmatic optical detection schemes, to see how the measurement incompatibility is manifested in practice. To do this we introduce the following figure of merit to quantify the measurement incompatibility for a specific detection scheme $O_k$, as illustrated in Fig.~\ref{scheme}, as follows:
\begin{equation}
    \mathcal{R}_{\{ O_k \}}(\rhol) = 1 - \frac{C^S(\rhol)}{F^{\text{(max)}}_\varphi \Delta^2 \varphi +  F^{\text{(max)}}_\eta \Delta^2 \eta},
    \label{measurement_incomp}
\end{equation}
where the variances are obtained from the particular measurement scheme, in our case photon counting or homodyne detection.

Furthermore, in order to consider a more general detection scheme, an additional beamsplitter with transmissivity $\tau_{\mathrm{out}} \in [0, 1]$ is placed before the detection, as shown in Fig.~\ref{scheme}, which transforms the output modes $\hat{b}_k$/$\hat{b}^\dagger_k$ into the detection modes $\hat{c}_k$/$\hat{c}^\dagger_k$, we have:
\begin{equation}
     \begin{pmatrix}
    \hat{c}_1 \\
    \hat{c}_2
    \end{pmatrix} 
    =
    \begin{pmatrix}
    \sqrt{\tau_{\mathrm{out}}} & -i \sqrt{1-\tau_{\mathrm{out}}}  \\
    -i \sqrt{1-\tau_{\mathrm{out}}} & \sqrt{\tau_{\mathrm{out}}}
    \end{pmatrix}
     \begin{pmatrix}
    \hat{b}_1 \\
    \hat{b}_2
    \end{pmatrix} .
    \label{input_beams}
\end{equation}
Let $O_1(\hat{c}_1, \hat{c}^\dagger_1)$ and $O_2(\hat{c}_2, \hat{c}^\dagger_2)$ be the two observables chosen for the measurement strategy in each output mode, as shown in Fig.~\ref{scheme}~(c).
The variances for the phase and loss are given by the generalized error propagation formula:
\begin{equation}
\Delta^2 \lambda_i (\rho_{\bm \lambda}) := \left[ 
\begin{pmatrix}
\frac{d  \langle O_1 \rangle}{d \lambda_i} & \frac{d  \langle O_2 \rangle }{d \lambda_i}
\end{pmatrix}
C_V^{-1}
\begin{pmatrix}
\frac{d  \langle O_1 \rangle}{d \lambda_i} \\ \frac{d  \langle O_2 \rangle}{d \lambda_i}
\end{pmatrix}
\right]^{-1} ,
\label{var}
\end{equation}
with the covariance matrix being defined as follows:
\begin{equation}
C_V = 
\begin{pmatrix}
 \text{Cov}(O_1, O_1)  & \text{Cov}(O_1, O_2) \\
\text{Cov}(O_2, O_1)  &  \text{Cov}(O_2, O_2) 
\end{pmatrix} ,
\end{equation}
where the expectation values are given by $\langle \cdot \rangle = \Tr \left( \hspace{1mm} \cdot \hspace{1mm}\rho_{\bm \lambda} \right)$ and the covariances are $\text{Cov}(O_i, O_j) = \langle O_i  O_j + O_j O_i \rangle/2 - \langle O_i \rangle \langle O_j \rangle$. 
Notice that here we consider the practical approach in which the parameters are estimated from the mean values of these observables.
Exploiting the full set of measurement outcomes, i.e. the projective measurement corresponding to the eigenstates of these observables, one 
may obtain more information about the parameters, as quantified by the classical Fisher information, at the expense of having to build a more complex estimator.

\subsubsection{\label{number} Photon counting}

The first detection scheme considered is the photon counting, by choosing $O_j = \hat{c}^\dagger_j \hat{c}_j$ in Eq.~\eqref{var}.
The calculations are indeed much simpler in the basis of total photon number $O_{+} = \hat{c}^\dagger_1 \hat{c}_1 + \hat{c}^\dagger_2 \hat{c}_2$ and photon number difference $O_{-} = \hat{c}^\dagger_1 \hat{c}_1 - \hat{c}^\dagger_2 \hat{c}_2$. Considering the two-mode Gaussian states, for the phase estimation we have the following derivatives: $\partial_\varphi \langle O_{+} \rangle = 0$ and $\partial_\varphi \langle O_{-} \rangle = 4 \sqrt{\eta \tau_{\mathrm{in}} (1-\tau_{\mathrm{in}}) \tau_{\mathrm{out}} (1-\tau_{\mathrm{out}})} \bar{N}_\alpha$; and for the loss estimation we have: $\partial_\eta \langle O_{+} \rangle = \tau_{\mathrm{in}} \bar{N}_\alpha + \bar{N}_r $ and $\partial_\eta \langle O_{-} \rangle = (2 \tau_{\mathrm{out}}-1) \left( \tau_{\mathrm{in}} \bar{N}_\alpha + \bar{N}_r \right)$. We start our analysis considering the state $| \psi^{(2)}_\alpha (0) \rangle$ with $\tau_{\mathrm{in}} = 1 - 1/\bar{N}^q$ and $\theta_2 =  2 \mu$. To achieve the SQL scaling in the phase estimation, the squeezing and displacement should be oriented in the opposite direction: $\theta_1 - 2 \mu = \pi$, as well as a balanced beamsplitter at the output: $\tau_{\mathrm{out}}=1/2$. With these considerations, the phase variance achieves the ultimate precision limit when $q<p$, as given by:
\begin{equation}
    \Delta^2 \varphi \left(
    \rho^{(2)}_\alpha (0) \right) F^{\text{(max)}}_\varphi \overset{N \rightarrow \infty}{\longrightarrow} 
    \begin{cases} 
     1 & \text{if } q < p, \\
     1 + \frac{1}{2 (1-\eta)} & \text{if } q = p, \\
     \infty & \text{if } q > p.
\end{cases}
\end{equation} 
and conversely. In contrast, to achieve the SQL scaling in the loss estimation, the squeezing and displacement should be oriented in the same direction: $\theta_1 - 2 \mu = 0$ for any value of $\tau_{\mathrm{out}}$, which is optimal at $\tau_{\mathrm{out}}=1$, maximizing both $\partial_\eta \langle O_{+} \rangle$ and $\partial_\eta \langle O_{-} \rangle$. In that case, we have:
\begin{equation}
   \Delta^2 \eta \left( 
    \rho^{(2)}_\alpha (0) \right) F^{\text{(max)}}_\eta \overset{N \rightarrow \infty}{\longrightarrow}
    \begin{cases} 
    1 + \frac{\eta}{1-\eta}
    & \text{if } q < p, \\
    1 + \frac{9 \eta}{10 (1 - \eta)}
    & \text{if } q = p, \\
    1 + \frac{\eta}{2 (1-\eta)}
    & \text{if } q > p.
\end{cases}
\end{equation}
which approaches the ultimate precision limit in the regime of large losses. Therefore, the measurement incompatibility arises from the different output transmissivity $\tau_{\mathrm{out}}$ required for each parameter, as also illustrated in Fig.~\ref{variances_vs_bound}~(a).
Additionally, for this measurement strategy there is also an incompatibility in the probe state, since a different orientation between the squeezing and displacement phase is required for each parameter. This probe incompatibility arises at the level of the QFI only when the input transmissivity $\tau_{\mathrm{in}}$ converges to one too fast ($q>p$), recovering the single-mode Gaussian state performance, according to Eqs.~\eqref{qfi_phase_sm} and \eqref{qfi_loss_sm}.

In Fig.~\ref{variances_vs_bound}(a),(b), we show the combined variances using the half photon strategy (i.e., half photons prepared in the optimal scheme for phase estimation and half for loss estimation).
Fig.~\ref{variances_vs_bound}(c) presents the independent estimation variances for phase and loss as functions of the output transmissivity $\tau_{\mathrm{out}}$, clearly illustrating the tradeoff in choosing the measurement: the phase variance is minimized when the loss variance is maximized, and vice versa.

We proceed by analyzing the state $| \psi^{(2)}_\alpha (\pi/2) \rangle$.
Now, we cannot set $\tau_{\mathrm{in}}=1$ as considered in the QFI, since it results in $\partial_\varphi \langle O_{+} \rangle = \partial_\varphi \langle  O_{-} \rangle  = 0$.
Therefore, to achieve the SQL for the phase estimation, we set $\tau_{\mathrm{out}}=1/2$, arriving at the following limit for the phase precision:
\begin{equation}
    \Delta^2 \varphi \left( 
    \rho^{(2)}_\alpha (\pi/2) )\right) F^{\text{(max)}}_\varphi \overset{N \rightarrow \infty}{\longrightarrow} 
    \frac{\eta + 1 - (1-\eta) \tau_{\mathrm{in}}}{(1-\eta) (1-\tau_{\mathrm{in}}) \tau_{\mathrm{in}}},
\end{equation}
which approaches the quantum CRB only when $\tau_{\mathrm{in}} \approx 1$ and $\eta \approx 0$.
In addition, from the previous equation, we conclude that even considering the input transmissivity in the form $\tau_{\mathrm{in}}=1-1/\bar{N}^q$ the phase variance diverges.
However, for loss estimation, this state cannot achieve the SQL scaling.
In order to achieve it, is necessary to choose an energy distribution in the form $\bar{N}_\alpha = k \bar{N}$ and  $\bar{N}_r = (1-k) \bar{N}$, in such a way that we achieve the SQL scaling only when $k=0$ (corresponding to a two-mode squeezed state, with $\Delta^2 \eta = 2 \eta (1-\eta)/\bar{N}$) or $k=1$ (corresponding to a coherent state, with $\Delta^2 \eta = \eta/\bar{N}$). Similarly to the previous state, Fig.~\ref{variances_vs_bound}(a),(b) shows the combined variances for the half photon strategy and in Fig.~\ref{variances_vs_bound}(c) the variances obtained from independent estimation in function of $\tau_{\mathrm{out}}$.

Finally, for the state $|\psi^{(2)}_N \rangle$, the combined variances obtained from the photon counting are shown in Fig.~\ref{variances_vs_bound}(a),(b), also considering the half photon strategy.
In Fig.~\ref{variances_vs_bound}(c) we show the independent estimation variances as a function of the output transmissivity, which  performs similarly to the Gaussian state $| \psi^{(2)}_{\alpha} (0) \rangle$. 

\subsubsection{\label{homodyne} Homodyne detection}

In previous works, homodyne detection has been demonstrated as a promising detection strategy for Gaussian states.
For instance, in the case of a bright two-mode squeezed state, it was shown that an improved homodyne detection saturates the corresponding quantum CRB for each parameter, when choosing the correct quadrature for each parameter~\cite{Dowran2021}.
Here, we consider conventional homodyne detection, by setting $O_j = e^{i \xi_j} \hat{c}^\dagger_j + h.c. := X_j(\xi_j)$ in Eq.~\eqref{var}, where we prove that the ultimate quantum precision limit can also be achieved asymptotically for each parameter; however, measurement incompatibility manifests itself in the fact that a different quadrature phase $\xi$ is optimal for each parameter.
This measurement strategy is implemented only for the Gaussian states, since the first moments of $X_j(\xi_j)$ vanish for the state with $N$ photons $| \psi^{(2)}_N \rangle$.

For two-mode Gaussian states, we have the following derivatives for phase estimation: $\partial_\varphi \langle X_1 (\chi) \rangle = 2 \sqrt{\eta  \tau_{\mathrm{in}} \tau_{\mathrm{out}}} \bar{N}_\alpha  \cos (\mu -\chi) $ and $\partial_\varphi \langle X_2 (\chi) \rangle = 2 \sqrt{\eta  \tau_{\mathrm{in}} (1-\tau_{\mathrm{out}})} \bar{N}_\alpha  \sin (\mu -\chi ) $.
For loss estimation we have:   $\partial_\eta \langle X_1 (\chi) \rangle = \sqrt{\tau_{\mathrm{in}} \tau_{\mathrm{out}}} \bar{N}_\alpha \sin (\mu -\chi ) / \sqrt{\eta}$ and $\partial_\eta \langle X_2 (\chi) \rangle = \sqrt{  \tau_{\mathrm{in}} (1-\tau_{\mathrm{out}})} \bar{N}_\alpha \cos (\mu -\chi ) / \sqrt{\eta}$. 

Let us start by considering the state $| \psi^{(2)}_\alpha (\pi/2) \rangle$ with phase parameters $\theta_1 - 2 \xi = \theta_2 - 2 \xi = 0, \pi$, and transmissivity $\tau_{\mathrm{in}}=1-1/N^p$.
With this choice, the variance achieves the SQL scaling when $\tau_{\mathrm{out}}=1$, with the following asymptotic expressions:
\begin{equation}
    \Delta^2 \varphi \left(\rho^{(2)}_\alpha (0) \right) F^{\text{(max)}}_\varphi \overset{N \rightarrow \infty}{\longrightarrow} 
    \begin{cases} 
     \sec^2(\mu + \xi) & \text{if } q < p , \\
     \big[ 1 + \frac{\eta}{2 (1-\eta)} \big] \sec^2(\mu + \xi) & \text{if } q = p , \\
     \infty & \text{if } q > p ,
\end{cases}
\end{equation}
and the limit for the loss precision is:
\begin{equation}
    \Delta^2 \eta \left( \rho^{(2)}_\alpha (0)\right) F^{\text{(max)}}_\eta \overset{N \rightarrow \infty}{\longrightarrow}
    \begin{cases} 
    \csc^2(\mu + \xi) & \text{if } q < p, \\
    \big[ 1 + \frac{\eta}{2 (1-\eta)} \big] \csc^2(\mu + \xi) & \text{if } q = p \, . \\
    \infty & \text{if } q > p,
\end{cases}
\end{equation}

We proceed by considering the state $| \psi^{(2)}_\alpha (\pi/2) \rangle$ with $\tau_{\mathrm{in}}=1$.
In this case, the variances achieve the SQL scaling when the output transmissivity is $\tau_{\mathrm{out}}=1$ and the squeezing phase is $\theta - 2 \xi = \pm \pi/2$, resulting in the asymptotic expression for the phase variance:
\begin{equation}
    \Delta^2 \varphi \left(\rho^{(2)}_\alpha (\pi/2)\right) F^{\text{(max)}}_\varphi \overset{N \rightarrow \infty}{\longrightarrow} \sec^2(\mu + \xi) ,
\end{equation}
and for the loss variance:
\begin{equation}
    \Delta^2 \eta \left(\rho^{(2)}_\alpha (\pi/2)\right) F^{\text{(max)}}_\varphi \overset{N \rightarrow \infty}{\longrightarrow} \csc^2(\mu + \xi), 
\end{equation}
Therefore, for both states, the ultimate precision limit for the phase estimation is asymptotically achieved when  $\mu - \xi = 0, \pi$, maximizing the phase signal $\partial_\varphi \langle X_1 (\xi) \rangle$, but suppressing the loss signal.
In contrast, for loss estimation, the optimal phase is  $\mu - \xi = \pm \pi/2$,  maximizing the loss signal $\partial_\eta \langle X_1 (\xi) \rangle$, but suppressing the phase signal. \me{We note that this incompatibility in choosing the optimal phase was previously manifested in the single-mode Gaussian states itself, as shown in Eqs.~\eqref{qfi_sm_alpha},~\eqref{qfi_phase_sm},~\eqref{qfi_loss_sm}, whereas here, it is avoided at the state level and instead appears in the measurement stage.}
Additionally, this incompatibility is illustrated in Fig.~\ref{variances_vs_bound}(c).
The combined variances for phase and loss estimation is show in Fig.~\ref{variances_vs_bound}(a),(b) considering a simultaneous estimation strategy with a balanced choice of the measured quadrature, i.e., $\mu + \xi = \pi/4$.

\section{\label{sec_conclusions}Conclusions}
We have provided a comprehensive perspective on the problem of simultaneous phase and loss estimation in optical interferometry. We have shown that  
probe-incompatibility may be overcome, either by a careful engineering of single mode non-Gaussian state, or utilizing mode entanglement with a lossless reference beam.
Our results provide strong evidence that measurement incompatibility cannot be overcome, even asymptotically, for states that can surpass probe incompatibility.
We have also carefully benchmarked all the results obtained with the fundamental bounds, as well as introduced powerful numerical optimization techniques that have not been used in previous studies of the topic. 

In this paper, we have not touched on the distinction between HCRB and tighter bounds that apply when only single-copy measurements are available, see e.g.~\cite{Conlon2020,Hayashi2023a}.
However, we mention that in scenario (ii) the HCRB can be saturated at the single-copy level~\cite{Albarelli2019,Conlon2020}, even though this may in general require a non-demolition measuremenet of the total number of photons.
As a perspective for future work, it would be interesting to
consider all incompatibility aspects in a unified manner.
Some mathematical and numerical tools to analyze this scenario have been introduced recently~\cite{Hayashi2024}, but it is unclear how to effectively employ them in the challenging, yet most interesting, regime of a large number of photons.

One of the possible further studies along these lines is the problem of simultaneous phase and loss estimation in presence of thermal noise (only studied for single-parameter phase~\cite{Gagatsos2017} or loss~\cite{Jonsson2022}), which may be relevant in microwave sensing regime, e.g. for quantum illumination tasks~\cite{Zhao2025}.

\section*{Acknowledgments}
M. E. O. B. acknowledges the useful discussions with Piotr Dulian regarding the implementation of the ISS algorithm.
M. E. O. B. acknowledges financial support by the São Paulo Research Foundation (FAPESP), Brasil, Process Number 
2022/13635-3.
F. A. acknowledges financial support from Marie Skłodowska-Curie Action EUHORIZON-MSCA-2021PF-01 (project QECANM, grant n. 101068347).
R. D. D. acknowledges support from the National Science Center (Poland) grant No.2020/37/B/ST2/02134.

\section*{Note}
This is the Accepted Manuscript version of an article accepted for publication in \emph{Journal of Physics A: Mathematical and Theoretical}. IOP Publishing Ltd is not responsible for any errors or omissions in this version of the manuscript or any version derived from it. The Version of Record is available online at \url{https://doi.org/10.1088/1751-8121/ade516}.

\appendix

\section*{Data availability}

The data that support the findings of this study are openly available at the following URL/DOI:~\url{https://github.com/Matheus-Eiji/incomp_phase_loss.git}

\section{\label{optimization} Implementation of the iterative see-saw algorithm}

In this section, it will be discussed in detail how the iterative see-saw algorithm is implemented for the optimization of the two states $| \psi^{(1)}_N \rangle$ and $| \psi^{(2)}_N \rangle$, considered in Sec. \ref{probe_incomp_nphotons}. To begin, the action of the quantum channel can be expressed in terms of the corresponding Kraus operators as follows 
\begin{equation}
    \rho_{\bm \lambda} = \Lambda_{\bm \lambda}(| \psi \rangle \langle \psi |) = \sum^N_{m=0} K_m \hspace{0.5mm} | \psi \rangle \langle \psi | \hspace{0.5mm} K^\dagger_m
    ,
    \label{expansion_alg}
\end{equation}
and additionally, the derivative with respect to the parameter $\lambda_i = \varphi, \eta$ is given by:
\begin{equation}
    \frac{\partial \rho_{\bm \lambda}}{\partial \lambda_i} = \sum^N_{m=0} \Bigg[ \Big( \frac{\partial K_m}{\partial \lambda_i} \Big) | \psi \rangle \langle \psi | \hspace{0.5mm} K^\dagger_m + K_m \hspace{0.5mm} | \psi \rangle \langle \psi | \hspace{0.5mm} \Big( \frac{ \partial K^\dagger_m}{\partial \lambda_i} \Big)  \Bigg] 
    \label{derivative_alg}
\end{equation}
The matrix $ M_{\lambda_i}$ in Eq.~\eqref{qfi_algorithm} can also be written in terms of the Kraus operators as follows:
\begin{equation}
    M_{\lambda_i} = \sum^N_{m=0} \Tr \Bigg[ 2 \Big(\frac{ \partial K^\dagger_m}{\partial \lambda_i} \Big) A K_m + 2 K^\dagger_m A \Big(\frac{ \partial K_m}{\partial \lambda_i} \Big) -  \sum^N_{m=0} K^\dagger_m A^2 K_m \Bigg]
    \label{matrixm_alg}
\end{equation}
Following, we define the matrix that imprints the phase shift and the effect of the loss in the probe state, respectively, as:
\begin{equation}
    U_\varphi = 
    \begin{pmatrix}
    1  & & & 0 \\
    & \rme^{i \varphi} & & \\
    & & \ddots & \\
    0 & & & \rme^{i N \varphi}
    \end{pmatrix}
    \qquad
    B_m = 
    \begin{pmatrix}
    \sqrt{b^m_m}  & & & 0 \\
    & \sqrt{b^{m+1}_m} & & \\
    & & \ddots & \\
    0 & & & \sqrt{b^N_m}
    \end{pmatrix} 
\end{equation}
which gives the corresponding Kraus operators for the two-mode and single-mode cases, as discussed in the following.

\subsection{Single-mode state}

To begin, we can write the single-mode state defined in Eq.~\eqref{psiN_sm} in the basis $\mathcal{B}^{(1)} = \{ | 0, 0 \rangle, | 1, 0 \rangle , ..., | N, 0 \rangle \}$ as the column vector $ | \psi^{(1)}_N \rangle = (c^{(1)}_0, c^{(1)}_1, ..., c^{(1)}_N)$.
Then, in that basis, the density matrix at the output in Eq.~\eqref{rho1_output} is given by:
\begin{equation}
    \rho^{(1)}_N = \sum^N_{m=0} 
    K^{(1)}_m 
    \begin{pmatrix}
        c^{(1)}_0 \\ \vdots \\ c^{(1)}_N
    \end{pmatrix}
    \Bigg[
     K^{(1)}_m 
    \begin{pmatrix}
        c^{(1)}_0 \\ \vdots \\ c^{(1)}_N
    \end{pmatrix}
    \Bigg]^\dagger 
    \label{kraus_sm_alg}
\end{equation}
where introducing $\widetilde{m} = N - m$, the Kraus operators matrices read as 
\begin{equation}
    K^{(1)}_m = 
    \begin{pmatrix}
    0_{\widetilde{m} \times m} & B_m \\
    0_{m \times m} & 0_{m \times \widetilde{m}}
    \end{pmatrix} 
    U_\varphi ,
\end{equation}
which give the state evolution in Eq.~\eqref{expansion_alg} for this single-mode state. Indeed, in Eq.~\eqref{rho1_output} the state with $m$ photons lost is given by $| \psi^{(1)}_m \rangle = K^{(1)}_m | \psi^{(1)}_N \rangle $. 

Finally, to implement the ISS algorithm described in Sec. \ref{probe_incomp_nphotons}, the only remaining step is to compute the derivatives of the Kraus operators, from which we obtain the corresponding SLDs $L_{\lambda_i}$ and the expression for the matrices $M_{\lambda_i}$. From the previous equation, the derivatives in the Kraus matrices give:
\begin{equation}
    \partial_{\lambda_i} K^{(1)}_m = \Gamma^{(1)}_{\lambda_i, m} K^{(1)}_m
    \label{derivatives_sm_alg}
\end{equation}
where, introducing $m \leq n \leq N$, we have for each parameter:
\begin{equation}
    \Gamma^{(1)}_{\varphi, m} = 
    \begin{pmatrix}
    & \ddots  & & \\
    0_{\widetilde{m} \times m} & & i n & \\
    & & & \ddots \\
    0_{m \times m} & & 0_{m \times \widetilde{m}} & 
    \end{pmatrix}
    \qquad
    \Gamma^{(1)}_{\eta, m} = 
    \begin{pmatrix}
    & \ddots  & & \\
    0_{\widetilde{m} \times m} & & \frac{n(1-\eta) - m}{2 \eta (1-\eta)} & \\
    & & & \ddots \\
    0_{m \times m} & & 0_{m \times \widetilde{m}} &
    \end{pmatrix}
    \label{gamma_sm}
\end{equation}
Remembering the definition of the SLDs $L_{\lambda_i}$ given in Eq.~\eqref{qfi}, we obtain it by replacing the previous derivatives in Eq.~\eqref{derivative_alg} and solving the linear equation $\partial_{\lambda_i} \rho_{\bm \lambda} =  \{ \rho_{\bm \lambda} , L_{\lambda_i} \}  / 2$.
Finally, the only remaining step is replace Eqs.~\eqref{kraus_sm_alg}),~\eqref{derivatives_sm_alg} in Eq.~\eqref{matrixm_alg} and then,  compute the eigenvectors and eigenvalues of the matrix $M_{\lambda_i}$, as described in Sec.~\ref{probe_incomp_nphotons}.

\subsection{Two-mode state}

Now, the two-mode state defined in Eq.~\eqref{psiN_tm} can be written in the basis with no photons lost, $\mathcal{B}^{(2)}_0 = \{ | 0, N \rangle, | 1, N-1 \rangle , ..., | N, 0 \rangle \}$, as the column vector $ | \psi^{(2)}_N \rangle = (c^{(2)}_0, c^{(2)}_1, ..., c^{(2)}_N)$.
At the output, according to Eq.~\eqref{rho2_output} the density matrix has a block structure where each block is the subspace with $m$ photons lost.
Then, in each block, we write our state in the basis with $m$ photons lost, $\mathcal{B}^{(2)}_m = \{ |n-m, N-n \rangle_{n \in [m,N]} \}$, obtaining the following density matrix at the output:
\begin{equation}
    \rho^{(2)}_N = \bigoplus^N_{m=0} 
    K^{(2)}_m 
    \begin{pmatrix}
        c^{(2)}_0 \\ \vdots \\ c^{(2)}_N
    \end{pmatrix}
    \Bigg[
     K^{(2)}_m 
    \begin{pmatrix}
        c^{(2)}_0 \\ \vdots \\ c^{(2)}_N
    \end{pmatrix}
    \Bigg]^\dagger 
    \label{kraus_tm_alg}
\end{equation}
where introducing $\widetilde{m} = N - m$, the Kraus operator matrices acting on each block are given by:
\begin{equation}
    K^{(2)}_m = 
    \begin{pmatrix}
    0_{\widetilde{m} \times m} & B_m \\
    \end{pmatrix} 
    U_\varphi ,
\end{equation}
which define the evolution given in Eq.~\eqref{expansion_alg} for this two-mode state.
Indeed, in Eq.~\eqref{rho2_output} the state with $m$ lost photons is given by $| \psi^{(2)}_m \rangle = K^{(2)}_m | \psi^{(2)}_N \rangle $. 

Similarly to the single-mode state, the derivatives of the Kraus matrices are given by:
\begin{equation}
    \partial_{\lambda_i} K^{(2)}_m = \Gamma^{(2)}_{\lambda_i, m} K^{(2)}_m
    \label{derivatives_tm_alg}
\end{equation}
where, introducing $m \leq n \leq N$, we have for each parameter:
\begin{equation}
    \Gamma^{(2)}_{\varphi, m} = 
    \begin{pmatrix}
    \ddots  & & \\
    &i n & \\
    & &  \ddots 
    \end{pmatrix}
    \qquad
    \Gamma^{(2)}_{\eta, m} =
    \begin{pmatrix}
         \ddots  & & \\
         & \frac{n(1-\eta) - m}{2 \eta (1-\eta)} & \\
         & &  \ddots 
    \end{pmatrix} .
    \label{gamma_tm}
\end{equation}
Additionally, for the two-mode state, the SLDs have a block structure in the same way as the density matrix \cite{Crowley2014},
\begin{equation}
    L^{(2)}_{\lambda_i} = \bigoplus^N_{m=0} L^{(2)}_{\lambda_i, m}.
\end{equation}
Each block corresponds to the SLD of the state with $m$ photons lost, which has an analytical solution~\cite{Crowley2014}, given in terms of the present notation by
\begin{equation}
    L^{(2)}_{\lambda_i, m} =\frac{2}{\Tr ( \rho^{(2)}_m )}  \left[ \Gamma^{(2)}_{\lambda_i, m} \rho^{(2)}_m + \rho^{(2)}_m (\Gamma^{(2)}_{\lambda_i, m})^* \right] - \frac{\rho^{(2)}_m}{[\Tr ( \rho^{(2)}_m )]^2} \Tr \left[ \left( \Gamma^{(2)}_{\lambda_i, m} + (\Gamma^{(2)}_{\lambda_i, m})^* \right) \rho^{(2)}_m \right]
    \label{sld_tm}
\end{equation}
where we denoted $\rho^{(2)}_m = | \psi^{(2)}_N \rangle \langle \psi^{(2)}_N |$.
Finally, the matrix $M_{\lambda_i}$ is obtained by replacing Eqs.~\eqref{kraus_tm_alg},~\eqref{derivatives_tm_alg} in Eq.~\eqref{matrixm_alg}, analogously to the single-mode state.

Indeed, Fig.~\ref{gaussian_vs_optimization} shows that the optimal normalized QFI for both single-mode and two-mode states converges for small losses ($\eta \approx 1$).
In that regime both states are approximated pure and the binomial terms in Eq.~\eqref{transform_phase_loss} become negligible when $m > 0$.
As a result, the Kraus expansions in Eq.~\eqref{expansion_alg} can be approximated $\Lambda_{\bm \lambda}(\cdot) \approx K_0 \hspace{0.5mm} \cdot \hspace{0.5mm} K^\dagger_0$ and the derivatives also converge since from Eqs.~\eqref{gamma_sm},~\eqref{gamma_tm} we have $\partial_{\lambda_i} K^{(1)}_0 = \partial_{\lambda_i} K^{(2)}_0$.
Furthermore, both states are approximated pure and the corresponding SLDs of both states also converge.
In fact, Eq.~\eqref{sld_tm} becomes valid also for the single-mode state. Consequently, the $M_{\lambda_i}$ matrices defined in Eq.~\eqref{matrixm_alg} are approximately the same for both states, resulting then in similar optimization outcomes $\{ 
c^{(1)}_n \}$ and $\{ c^{(2)}_n \}$. This explains why the optimization yields nearly identical normalized QFI for $\eta=0.9$ at Fig.~\ref{gaussian_vs_optimization}.

\section{\label{gaussian_calculations} Calculation of the QFI matrix and the HCRB for the Gaussian states}

For a Gaussian state with covariance matrix $\sigma$ and displacement ${\bf d}$, the elements of the quantum Fisher information matrix are given by \cite{Safranek2019}:
\begin{equation}
F_{ij} = \lim_{\nu \rightarrow 1} \frac{1}{2} \text{vec}(\partial_{\lambda_i} \sigma )^\dagger M^{-1} \text{vec}(\partial_{\lambda_j} \sigma ) + 2 (\partial_{\lambda_i} {\bf d} )^\dagger \sigma^{-1} (\partial_{\lambda_j} {\bf d})
\label{qfi_general_gaussian}
\end{equation}
where we introduce the matrix $M = \nu^2 \sigma^* \otimes \sigma - \Omega \otimes \Omega $, $\Omega =  I_2 \oplus (-I_2)$ is the symplectic form, $\text{vec}(A)$ the vectorization of the matrix $A$.
Another important quantity is the expectation value of the commutator of two SLDs $\hat{L}_i$ and $\hat{L}_j$, which is given by \cite{Safranek2019}:
\begin{eqnarray}
\Tr \left( \hat{\rho}[\hat{L}_{\lambda_i},\hat{L}_{\lambda_j}] \right)  &=& 4 (\partial_{\lambda_i} {\bf d} )^\dagger \sigma^{-1} \Omega  \sigma^{-1} (\partial_{\lambda_j} {\bf d})  + \nonumber\\ && +  \text{vec}(\partial_{\lambda_i} \sigma )^\dagger M^{-1} (  \sigma^* \otimes \Omega - \Omega \otimes \sigma) M^{-1}  \text{vec}(\partial_{\lambda_j} \sigma )
\nonumber\\
\end{eqnarray}
From the last two equations, the single-mode state given in Eq.~\eqref{gaussian_singlemode} results in Eqs.~\eqref{qfi_sm_alpha}--\eqref{qfi_sm_correlation} in Sec.~\ref{section_gauss_sm}.
Additionally, the two-mode state given in Eq.~\eqref{gaussian_singlemode} results in the Eqs.~\eqref{qfi_phase_sm}--\eqref{qfi_correlation_tm}, in Sec.~\ref{tm}.

\me{Here, we focus on two-mode Gaussian states that can attain the ultimate precision limit for both parameters. Assuming the regime of strong displacement, we retain only the displacement contribution in Eq.~\eqref{qfi_general_gaussian}.
Under these conditions, from Eqs.~\eqref{cov_twomode}, ~\eqref{displacement_vector},~\eqref{qfi_general_gaussian} and considering $\bar{N}_r \gg 1$, the phase and loss QFIs for the state $| \psi^{(2)}_\alpha (0) \rangle$ are given, respectively, by the following expressions:
\begin{align}
F_{\varphi \varphi} ( \rho^{(2)}_\alpha (0) )  & \approx \frac{ 4 \eta \tau_{in} \Big[ 1 + \gamma (1-\eta ) (1-\tau_{in}) \tau_{in}  \bar{N}^2_r \Big] \bar{N}_\alpha
    }{
    (1-\eta ) \Big[ \gamma (1-\eta ) (1-\tau_{in}) \tau_{in}  \bar{N}^2_r + 4 \eta  \bar{N}_r \Big]+1
    } + \nonumber \\ 
& + \frac{8 \eta^2 \tau_{in} \Big[ 1 -
    \tau_{in} \cos (\theta_1-2 \mu )+(1-\tau_{in}) \cos(\theta_2-2 \mu )
    \Big] \bar{N}_r \bar{N}_\alpha
    }{ 
    (1-\eta ) \Big[ \gamma (1-\eta ) (1-\tau_{in}) \tau_{in}  \bar{N}^2_r + 4 \eta  \bar{N}_r \Big]+1
    } ,
\end{align}
\begin{align}
F_{\eta \eta} ( \rho^{(2)}_\alpha (0) ) & \approx \frac{ \tau_{in} \Big[ 1 +
    \gamma (1-\eta ) (1-\tau_{in}) \tau_{in}  \bar{N}^2_r \Big] \bar{N}_\alpha
    }{
    \eta (1-\eta ) \Big[ \gamma (1-\eta ) (1-\tau_{in}) \tau_{in}   \bar{N}^2_r+4 \eta  \bar{N}_r \Big]+\eta 
    } + \nonumber \\ 
& + \frac{
    2 \eta \Big[1 +\tau_{in} \cos (\theta_1-2 \mu )-(1-\tau_{in}) \cos (\theta_2-2 \mu ) \Big] \bar{N}_r \bar{N}_\alpha
    }{
    \eta (1-\eta ) \Big[ \gamma (1-\eta ) (1-\tau_{in}) \tau_{in}   \bar{N}^2_r+4 \eta  \bar{N}_r \Big]+\eta 
    } .
\end{align}
where we introduce the coefficient $\gamma = 16 \cos ^2\left(\frac{\theta_1-\theta_2}{2}\right)$. Similarly, the phase and loss QFIs for the state $| \psi^{(2)}_\alpha (\pi/2) \rangle$ are given, respectively, by the following expressions:
\begin{align}
F_{\varphi \varphi} ( \rho^{(2)}_\alpha (\pi/2) )  & \approx \frac{
    4 \eta  \tau_{in} \Big[ 1 + 4 (1-\eta ) (2 \tau_{in}-1)^2 \bar{N}^2_r \Big] \bar{N}_\alpha
    }{
    1 + 4 (1-\eta ) \Big[ 1 -4 (1-\eta ) (1-\tau_{in}) \tau_{in}+(1-\eta ) (1-2 \tau_{in})^2 \bar{N}_r \Big] \bar{N}_r
    } + \nonumber \\ 
& + \frac{
    8 \eta  \tau_{in} \Big[2 -\eta +2 \eta  \sqrt{(1-\tau_{in}) \tau_{in}} \sin (\theta -2 \mu )-8 (1-\eta ) (1-\tau_{in}) \tau_{in} \Big]  \bar{N}_r \bar{N}_\alpha
    }{
    1 + 4 (1-\eta ) \Big[ 1 -4 (1-\eta ) (1-\tau_{in}) \tau_{in}+(1-\eta ) (1-2 \tau_{in})^2 \bar{N}_r \Big] \bar{N}_r
    } ,
\end{align}
\begin{align}
F_{\eta \eta} ( \rho^{(2)}_\alpha (\pi/2) ) & \approx \frac{
    \tau_{in} \Big[ 1-4 \eta  \sqrt{(1-\tau_{in}) \tau_{in}} \sin (\theta -2 \mu ) \bar{N}_r\Big] \bar{N}_\alpha
    }{
    \eta +4 (1-\eta ) \eta  \Big[1-4 (1-\eta ) (1-\tau_{in}) \tau_{in}+(1-\eta ) (1-2 \tau_{in})^2 \bar{N}_r\Big] \bar{N}_r 
    } + \nonumber \\ 
& + \frac{
    4 \tau_{in} (1-\eta) (1-2 \tau_{in})^2 \bar{N}^2_r  \bar{N}_\alpha
    }{
    \eta +4 (1-\eta ) \eta  \Big[1-4 (1-\eta ) (1-\tau_{in}) \tau_{in}+(1-\eta ) (1-2 \tau_{in})^2 \bar{N}_r\Big] \bar{N}_r 
    } .
\end{align}
}

\begin{figure}[t]
\includegraphics[width=0.9 \columnwidth]{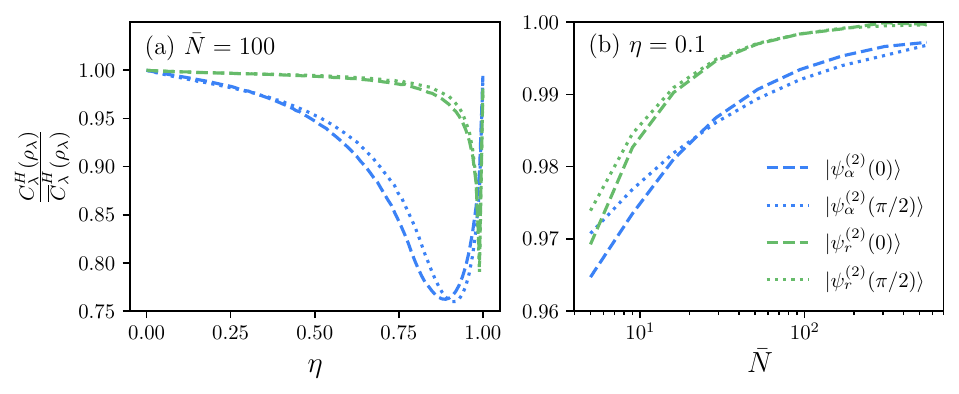}
\caption{Comparison of the HCRB, given in Eq.~\eqref{hcrb} and the upper bound for the HCRB, given in Eq.~\eqref{upper_hcrb},  for the two-mode Gaussian states considered.}
\label{hcrb_vs_upperhcrb}
\end{figure}

Recently, a method for numerically evaluating the HCRB for arbitrary Gaussian states based on the covariance matrix $\sigma$ and the displacement vector $\mathbf{d}$ was introduced~\cite{Chang2025}.
Following this approach, we compute the HCRB $C^H_{\bm \lambda}$ for two-mode Gaussian states and compare it with the upper bound $\overline{C}^H_{\bm \lambda}$, as shown in Fig.~\ref{hcrb_vs_upperhcrb}.

\section{\me{Probe incompatibility of two-mode states from minimization over Kraus representations}} \label{app:KrausMin}

We briefly recall the results of~\cite{Albarelli2022} for the numerical calculation of the bound~\eqref{eq:probeincompbound}, which amounts to solving the following optimization
\begin{equation}
    \label{eq:minKraus}
\min_{\{\tilde{K}_{\lambda,m}\}}  \left\| \sum_{j=1}^d \frac{1}{ w_{\lambda_j}} \sum_{m} \partial_{\lambda_j} \tilde{K}_{m}^\dagger \partial_{\lambda_j} \tilde{K}_{m} \right\|.
\end{equation}
In particular, to evaluate the probe incompatibility quantity we choose the weights $w_{\lambda_j} = F_{\lambda_j}^{\mathrm{(max)}}$, i.e. the optimal single-parameter QFI for each parameter.
We collect the Kraus operators in a column vector $\boldsymbol{K} = [ K_1, \dots K_N ]^T$, so that the different Kraus representations of the channel, related by an isometry as in Eq.~\eqref{eq:Kraus_equiv}, can be written in vector form as $\tilde{\boldsymbol{K}}= u(\bm \lambda) \boldsymbol{K}$.
The inner sum appearing in the bound can then be written as
\begin{equation}
\sum_{m} \partial_{\lambda_j} K_{m}^\dagger \partial_{\lambda_j} K_{m} = \partial_{\lambda_j} \tilde{\boldsymbol{K}}^\dag  \partial_{\lambda_j} \tilde{\boldsymbol{K} } =  
 \left( \partial_{\lambda_j} \boldsymbol{K} -i h_{\lambda_j} \boldsymbol{K} \right)^\dag \left( \partial_{\lambda_j} \boldsymbol{K} -i h_{\lambda_j} \boldsymbol{K}\right),
\end{equation}
where the dependence on the specific Kraus representation only appears in the choice of $d$ hermitian matrices $h_{\lambda_j} = i u(\bm \lambda)^\dag \partial_{\lambda_j} u(\bm \lambda)$, which have to be optimized.
By introducing matrix
\begin{equation}
\boldsymbol{D} = \begin{bmatrix}
\sqrt{1/w_{\lambda_1}} \left( \partial_{\lambda_1} \boldsymbol{K} - i h_{\lambda_1} \boldsymbol{K} \right) \\
\vdots \\
\sqrt{1/w_{\lambda_{d}}} \left( \partial_{\lambda_d} \boldsymbol{K} - i h_{\lambda_d} \boldsymbol{K} \right)
\end{bmatrix},
\end{equation}
The minimization in Eq.~\eqref{eq:minKraus} can be rewritten as the following SDP
\begin{equation}
\label{eq:SDPtotalQFI}
 \min_{t,\{ h_{\lambda_x} \}_{x=1}^d} t \\
\quad \quad \text{subject to }
\begin{bmatrix}
t \mathbb{1} & \boldsymbol{D}^\dag \\
\boldsymbol{D} & \mathbb{1}
\end{bmatrix} \geq 0.
\end{equation}

\begin{figure}[t]
    \centering
\includegraphics[width=.6 \columnwidth]{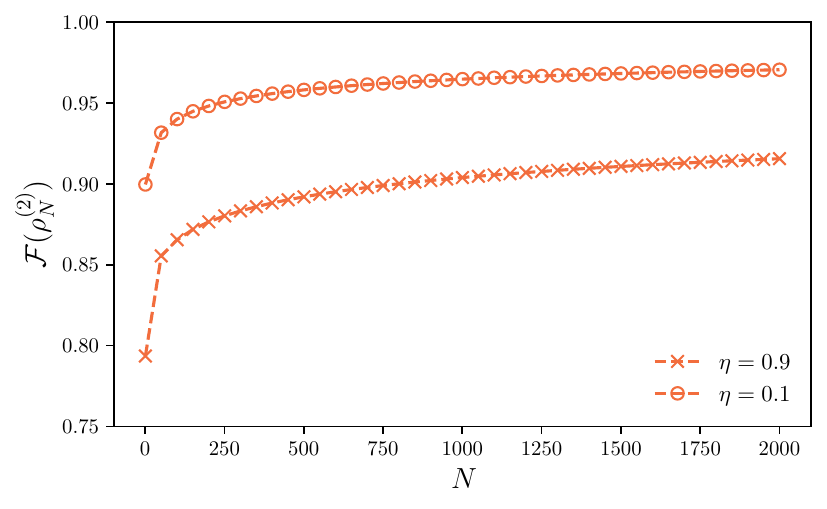}
\caption{Normalized QFI for optimal two-mode $N$-photons probe states, as in panels (b) and (c) of Fig.~\ref{gaussian_vs_optimization}. 
This quantity is here computed by minimizing over equivalent Kraus representations, allowing to reach $N = 2000$.
\label{fig:purif}
}
\end{figure}

To apply this technique to simultaneous estimation of phase and loss, we need the Kraus operators in Eq.~\eqref{eq:KrausOps_alpha_beta}\footnote{Reported here for $\alpha=\beta=0$, which correspond to a particular choice of the matrix $u(\bm \lambda)$.}
\begin{equation}
K_{ m} = \sqrt{\frac{(1-\eta)^m}{m!}}e^{i \varphi \hat{n}}\eta^{\frac{n}{2}} \hat{b}^m, \quad m=0,1,\dots, N,
\end{equation}
where we have fixed the maximal number of photons to $N$.
Following the same argument of~\cite[App.~G]{Zhou2020} the optimization can be restricted to matrices $h_{h_{\lambda_x}}$ thare are diagonal, so that all the operators involved in the minimization are diagonal in the Fock basis, making a numerical evaluation more efficient. 
More formally, the problem becomes a quadratically constrained quadratic program, instead of a fully-fledged SDP, but concretely we have used an SDP solver.
Numerical results for two exemplary values $\eta=0.1$ and $\eta=0.9$ are reported in Fig.~\ref{fig:purif}, where figures of merit up to $N=2000$ are shown, compared to the maximal value $N=1000$ in Fig.~\ref{gaussian_vs_optimization}.

We conclude this brief discussion by mentioning that this method does not immediately provide the optimal state $\rho^{(2)}_N$.
However, it can be obtained by solving a second SDP, see~\cite[App.~F]{Zhou2020} for the single-parameter case and~\cite[App.~G]{Albarelli2022} for the multiparameter case.
We have not pursued this approach in this work, and only used the ISS method to find optimal states.


\nocite{apsrev4Control}
\bibliography{biblio,revtex-custom}

\end{document}